\documentclass[aps,prd,eqsecnum,showpacs,amsmath,nofootinbib,superscriptaddress,twocolumn,floats,preprintnumbers]{revtex4}

\usepackage[dvips]{color,graphicx}
\usepackage{amsfonts,amssymb,theorem,mathrsfs,times}
\definecolor{red  }{rgb}{1,0,0}
\definecolor{blue }{rgb}{0,0,1}
\definecolor{green}{rgb}{0,1,0}

\newcommand{\bea}{\begin{eqnarray}}
\newcommand{\ena}{\end{eqnarray}}

\newcommand{\mn}{{\mu\nu}}

\renewcommand{\a}{\alpha}
\newcommand{\ta}{\tilde{\a}}

\renewcommand{\c}{\gamma}
\renewcommand{\d}{\delta}
\newcommand{\e}{\epsilon}
\renewcommand{\k}{\kappa}
\newcommand{\s}{\sigma}

\newcommand{\la}{\lambda}

\newcommand{\pa}{\partial}
\newcommand{\na}{\nabla}

\newcommand{\nn}{\nonumber \\}
\newcommand{\lh}{\left(}
\newcommand{\rh}{\right)}

\newcommand{\qq}{\qquad}
\newcommand{\eq}[1]{(\ref{#1})}
\newcommand{\vect}[1]{\!\!\!\mbox{ \boldmath $#1$}}
\newcommand{\E}{{\scriptscriptstyle E}}

\newcommand{\dalm}{\kern1pt\vbox{rule height 0.9pt\hat box{\vrule width
0.9pt\hat skip 2.5pt\vbox{\vskip 5.5pt}\hat skip 3pt\vrule width 0.3pt}rule height
0.3pt}\kern1pt}

\usepackage{amsmath}	
\begin{document}
\preprint{WU-AP/308/10}
\preprint{KU-TP 049}

\title{AdS Black Hole Solutions in Dilatonic Einstein-Gauss-Bonnet Gravity
}
\author{Kei-ichi Maeda}
\email{maeda"at"waseda.jp}
\affiliation{
Department of Physics, Waseda University,
Shinjuku, Tokyo 169-8555, Japan
}
\affiliation{
Advanced Research Institute for Science and Engineering,
Waseda University, Shinjuku, Tokyo 169-8555, Japan
}
\author{Nobuyoshi Ohta}
\email{ohtan"at"phys.kindai.ac.jp}
\affiliation{
Department of Physics, Kinki University, Higashi-Osaka, Osaka 577-8502,
Japan
}
\author{Yukinori Sasagawa}
\email{yukinori"at"gravity.phys.waseda.ac.jp}
\affiliation{
Department of Physics, Waseda University,
Shinjuku, Tokyo 169-8555, Japan
}

\date{\today}

\begin{abstract}
We find that anti-de Sitter (AdS) spacetime with a nontrivial linear dilaton field
is an exact solution in the effective action of the string theory,
which is described by gravity with the Gauss-Bonnet curvature terms
coupled to a dilaton field in the string frame without a cosmological constant.
The AdS radius is determined by the spacetime dimensions
and the coupling constants of curvature corrections.
We also construct the asymptotically AdS
black hole solutions with a linear dilaton field numerically.
We find these AdS black holes for hyperbolic topology and in dimensions higher than four.
We discuss the thermodynamical properties of those solutions.
Extending the model to the case with the even-order higher Lovelock curvature terms,
we also find the exact AdS spacetime with a nontrivial dilaton.
We further find a cosmological solution with a bounce of three-dimensional space
and a solitonic solution with a nontrivial dilaton field,
which is regular everywhere and approaches an asymptotically AdS spacetime.
\end{abstract}

\pacs{
}

\maketitle

\section{Introduction}

The so-called ``Lovelock gravity" \cite{Lovelock} is
a natural higher-curvature generalization of the Einstein gravity.
It includes the Einstein-Hilbert action
as the first Lovelock term, and the second term
is known as the Gauss-Bonnet (GB) term.
The minimum generalization of the Einstein gravity
is the Einstein-Gauss-Bonnet (EGB) gravity theory,
which contains up to the second Lovelock term.
Since higher-curvature terms may come from
quantum corrections of gravity,
the EGB gravity theory has been studied extensively.
Black hole solutions in the EGB gravity theory
were discussed in \cite{BH_EGB,dSEDEGB,AdSBH_EGB}.
The qualitative difference from the Einstein gravity is
the existence of branches of the solutions; i.e., not only
Minkowski spacetime but also anti-de Sitter (AdS) spacetime
exists as a vacuum state in the EGB gravity theory.
As a result, the black hole solutions also possess
both an asymptotically flat and AdS branches without a
negative cosmological constant.

In the context of the string theory,
the EGB gravity has been argued to be an effective field theory with
quantum corrections in \cite{Zumino}.
The GB term is indeed found as higher-order correction in string theories~\cite{MT}.
It is also necessary to include a dilaton field as well as higher-curvature
correction terms since the dilaton field plays an important role in the string theory.
(The vacuum expectation value of the dilaton field gives the string coupling constant.)
In a dilatonic gravity theory, we should first choose a ``frame''
where we discuss and interpret the physical meanings.
There are two fundamental frames:
One is a string frame where a dilaton coupling
is factored out in all terms in the action,
while the other is the Einstein frame where
the scalar-curvature term is minimally coupled to a dilaton field.
The string frame is the frame which is naturally defined
in the string theory, while the Einstein frame gives the Einstein gravity,
which is familiar to us, with higher-order corrections.

The gravity theory in the string frame and
that in the Einstein frame are classically equivalent
since we can transform from one frame to the other by
an appropriate conformal transformation~\cite{maeda_ohta_sasagawa}.
However, there appear extra higher-order terms via a conformal transformation
in addition to the gravity action and the conventional kinetic term of
a dilaton field, and those terms are sometimes ignored.
As a result, the two theories in these frames are different
when we ignore such extra terms.
We call those in the string frame and in the Einstein frame
DEGB (dilatonic Einstein-Gauss-Bonnet) gravity and EDEGB
(``Einstein-frame" DEGB) gravity, respectively.
The effective action in the Einstein frame is obtained from the scattering
amplitudes of gravitons and dilatons,
but that in the string frame is calculated by the beta function method.
Thus, in the approximation keeping terms up to certain order in the higher
derivatives, there are always ambiguities in the effective action as represented
by the freedom of field redefinition. Both of them have their own rights,
and it is significant to study both and check if they give qualitatively
different results.

Most of the works on dilatonic Einstein-Gauss-Bonnet gravity
have considered the EDEGB gravity theory.
Black hole solutions in the EDEGB gravity were studied for an asymptotically
flat case in four dimensions~\cite{degb,degb2,Torii_Yajima_Maeda} and
higher dimensions~\cite{ohta_torii1}.
The extremal black hole solutions, which are important in relation
with Bogomol'nyi-Prasad-Sommerfeld states in the supersymmetric theory, were studied in the EDEGB gravity
in four dimensions~\cite{Chen1,Chen2} and higher dimensions~\cite{Chen3}.
The extremal black hole solutions in the DEGB gravity were
also studied in~\cite{Davydov}.
The difference between the black holes in the EDEGB and the DEGB gravity theories
was analyzed for asymptotically flat
black hole solutions in~\cite{maeda_ohta_sasagawa}.
The relation between the EDEGB and the DEGB is discussed
from the viewpoint of Einstein's equivalence principle in \cite{iihoshi}.

AdS spacetime is attracting increasing attention since
the AdS/conformal field theory (AdS/CFT) correspondence was proposed.
According to the AdS/CFT, one may examine the strong coupling region of the boundary CFT
by considering the corresponding black hole solutions in bulk AdS spacetime.
In fact, black hole solutions in the EDEGB were studied in \cite{Cai1}, and
the quantum corrections to the ratio of
the shear viscosity to entropy density in the CFT side were calculated.
We expect that our solutions may also be useful in this kind of study.

An AdS black hole, which corresponds to the finite temperature CFT according to the AdS/CFT
correspondence, is known to possess interesting thermodynamical properties.
When the horizon topology is a sphere ($k=1$), the small AdS Schwarzschild black hole
is thermodynamically unstable while the large one is stable.
There exists a phase transition, namely, Hawking-Page transition~\cite{HawkingPage},
between the large AdS black hole and a thermal AdS space.
In contrast, 
planar ($k=0$) and hyperbolic ($k=-1$) AdS black holes are stable~\cite{Birmingham}
and the Hawking-Page transition does not occur.

In the EGB model with a negative cosmological constant, the thermodynamical properties
of the asymptotically AdS black hole solutions (``general relativity branch'') were studied
in \cite{AdSBH_EGB} and with a special emphasis on the hyperbolic case in \cite{Neupane}.
In the presence of a negative cosmological constant \cite{AdSBH_EGB},
thermodynamic properties of asymptotically AdS black hole solutions for $k=1$
are similar to those without the Gauss-Bonnet term in dimensions other than five.
In five dimensions there appears a new phase of stable small black holes.
For $k = 0$, black hole solutions are thermodynamically stable,
and for $k=-1$, black hole solutions are always stable.

The asymptotically AdS spacetimes in the EDEGB model were analysed
with and without the cosmological
constant~\cite{dSEDEGB,ohta_torii2,ohta_torii3,ohta_torii4},
and their global structures were studied in \cite{ohta_torii5}.
In this paper, we consider black holes and other exact solutions
in such spacetimes with a nontrivial dilaton field whose behavior is given by
$\phi \propto \ln r $ and discuss their thermodynamic properties.
This type of dilaton field is called a linear dilaton,
and it is essential for the existence of black hole solutions
in the two-dimensional dilatonic gravity model~\cite{FHL}.
In the Einstein-Maxwell-dilaton gravity, linear dilaton
solutions whose asymptotic behaviours are not
AdS nor de Sitter were studied in \cite{cai_wang}.
Some of such solutions are obtained from asymptotically
AdS or de Sitter solutions with a suitable compactification.
Intersecting brane solutions with linear dilatons are also constructed
in \cite{CGO}.

We find that AdS black hole solutions for $k=-1$ as well as
AdS spacetime for $k=0$ exist without a negative
cosmological constant if the dilaton is nontrivial, in contrast to,
for example, Refs.~\cite{ohta_torii2,ohta_torii3,ohta_torii4}.
The AdS black hole solutions are obtained numerically,
and the latter spacetime solutions are shown to exist for even-order
Lovelock gravity.
We also find interesting cosmological and solitonic solutions.

This paper is organized as follows.
In the next section, we introduce the DEGB gravity theory
in the string frame and show that 
the AdS spacetime is the exact solution in a plane symmetric case.
Then, in Sec.~III, we show how to find asymptotically AdS
black hole solutions in the DEGB theory. In Sec.~IV, we present our numerical
results and discuss the thermodynamical properties of the black holes.
Our conclusions are drawn in Sec.~V. In the appendices, we give an AdS solution
in the Lovelock theory of gravity and briefly discuss
a horizonless solitonic solution and
a cosmological solution obtained in our system.\\

\section{Dilatonic Einstein-Gauss-Bonnet gravity: basic equations}

The effective action of the heterotic string theory
in the string frame is given by
\begin{align}
{\cal S}_{\rm S} &=\frac{1}{2\k_{D}^2} \int d^D \!x \sqrt{-g} \ e^{-2\phi}
 \biggl(R +4(\na \phi)^2
 + \a_{2} R_{GB}^2 \biggr)
\,,
\label{Saction}
\end{align}
where $\k_{D}^2$ is the $D$-dimensional gravitational constant,
$\phi$ is a dilaton field,\footnotemark
\begin{align}
R_{GB}^2 := R^2 -4 R_{\mn} R^{\mn}
+R_{\mn\rho\s}R^{\mn\rho\s}
\end{align}
is the Gauss-Bonnet curvature term, and $\a_2=\alpha'/8$ is its coupling constant
with $\a'$ being the Regge slope.
First we present the basic equations.
To find a black hole solution, we assume a symmetric (depending only on $r$) 
and static spacetime, whose metric form is given by
\begin{align}
ds^2_D=-e^{2\nu}dt^2+e^{2\la}dr^2+e^{2\mu} d\Sigma_{k}^2\,,
\label{metric_string}
\end{align}
where $\nu,\la,$ and $\mu$ are functions of the radial coordinate $r$
and
\begin{align}
d\Sigma_{k}^2=\c_{ij}^{(k)} dx^i dx^j
\,,
\label{max_sym}
\end{align}
is the metric of $(D-2)$-dimensional maximally
symmetric hypersurface with a constant curvature of the sign $k=0,\pm 1$.

\begin{widetext}
First we show the explicit form of the action with this ansatz:
\begin{align}
{\cal S}_{\!S}=
\frac{1}{2\k_{D}^2} \int d^D \!x e^{W-2\phi}
 \biggl\{ & e^{-2\la}\bigl[ 2 (D-2) Y +(D-2)_3
A -4 \phi' \nu' \bigr]
 \biggr.
 +4 e^{-2\la} \phi'^2 \nn
 + & \ta_2 e^{-4\la}
 \bigl[ (D-4)_5 A^{2} +4 (D-4) Y A
 -8 \phi' \nu' A
\bigr] \biggl. \biggr\},
\end{align}
\end{widetext}
where we have introduced
\begin{eqnarray}
W&:=& ~ \nu +\la +(D-2)\mu \,,
\nonumber \\
Y&:=& ~ -\left(\mu''+\mu'^2 -\mu' \la'\right) \,,
\nonumber \\
A&:=& ~ k e^{ 2 ( \la -\mu )} -\mu'^2 \,,
\label{YAW} \nn
(D-m)_n&:=& (D-m)(D-m-1)\cdots (D-n)\,,~~
\end{eqnarray}
with $m$ and $n$ being integers ($n>m$)
and dropped the surface term. \footnotetext{The string theory is constructed in ten dimensions.
But we discuss arbitrary dimensions.}
A prime denotes a derivative with respect to $r$.
For brevity, we have introduced the rescaled coupling
constant $\ta_2 := (D-2)_3 \a_2$, and in what follows,
we will normalize the variables by $\ta_2$.
Taking variations of the action
with respect to $\phi, \ \nu, \ \la,$ and $\mu$,
we find the basic equations.
Since we are interested in a black hole solution,
it is convenient to introduce new metric functions
$f$ and $\delta$ as
\begin{gather}
ds^2 = -f(r) e^{-2 \d(r)} dt^2
+\frac{1}{ f(r)} dr^2 +r^2 d\Sigma_{k}^2,
\label{metric_string2}
\end{gather}
Here we have fixed one metric component as
$e^{2\mu}=r^2$ by using the gauge freedom of the radial coordinate $r$.
Using the new metric functions and defining the following variables by
\begin{eqnarray}
h&:=& r(f'-2 f \delta')
\,,
\nn
X&:=& \frac{1}{4 f^2
r^2} \left[ h(f'r-h) -2 f (h'
r - h ) \right]
\,,
\end{eqnarray}
the basic equations are written as
\begin{widetext}
\begin{align}
f r^4
 F_{{\rm S}(\phi)}&:= -2\Bigl\{ r^2 \bigl[ (D-2)_{3} (k-f) -(D-2)
(f'r+h) +2 X f r^2
 +2 (f'r+h) \phi' r +4(D-2) f \phi'r + 4 f ( \phi'' - \phi'^2 )
r^2 \bigr] \nn
&\hspace{0.7cm}+ \ta_2
 \bigl[
(D-4)_{5} (k-f)^2 -2 (D-4) (k-f) (f'r+h) +4
X (k-f) f r^2
+ 2 h f' r
 \bigr] \Bigr\} =0,
\label{eq_phi}\\
f r^4 F_{{\rm S}(\nu)}&:=
r^2 \bigl[(D-2)_3 (k-f)
 -(D-2) (f' - 4 f \phi' ) r +4 f ( \phi'' -\phi'^2 ) r^2
 +2 \phi' f' r^2 \bigr] \nn
&+ \ta_2
 \bigl[
(D-4)_{5} (k-f)^2 -2 (D-4) (k-f) (f' - 4 f \phi' ) r
+8 f (k-f) ( \phi'' -2 \phi'^2 ) r^2 +4 (k-3 f) \phi' f' r^2
 \bigr] =0, \label{eq_nu} \\
f r^4 F_{{\rm S}(\la)}&:=
r^2 \bigl[(D-2)_3 (k-f) -(D-2) ( h -4 f \phi' r )
 +2 ( h -2 f \phi' r ) \phi' r \bigr] \nn
&+ \ta_2
 \bigl[
(D-4)_{5} (k-f)^2 -2 (D-4) (k-f) ( h -4 f \phi' r )
 +4 (k-3 f) h \phi' r
 \bigr] =0,
\label{eq_lam}\\
f r^4 F_{{\rm S}(\mu)}&:=
r^2 \bigl[ (D-2)_4 (k-f) -(D-2)_3 (f'r+h) +4 (D-2)_3 f \phi' r
 +2 (D-2) X f r^2 \nn
& \hspace{1.6cm} +4(D-2) f ( \phi'' - \phi'^2 ) r^2
 + 2 (D-2) (f'r+h) \phi' r \bigr] \nn
&+ \ta_2
 \bigl[
(D-4)_{6} (k-f)^2 -2 (D-4)_5 (k-f) (f'r+h) +8 (D-4)_5 (k-f)
f \phi' r
\nn
& \hspace{1.6cm}
 +4(D-4) X (k-f) f r^2
+8(D-4) f (k-f) ( \phi'' - 2 \phi'^2 ) r^2
 +4(D-4) (k-3 f) (f'r+h) \phi' r \nn
& \hspace{1.6cm} +2 (D-4) h f' r -8 f h ( \phi'' - 2 \phi'^2 ) r^2
 -8 h f' \phi' r^2 +16 X \phi' f^2 r^3
 \bigr] =0
\,. \label{eq_mu}
\end{align}
\end{widetext}
From the Bianchi identity, we find one relation between the above
four functionals $ F_{{\rm S}(\phi)}, F_{{\rm S}(\nu)},
 F_{{\rm S}(\la)}$, and $ F_{{\rm S}(\mu)}$, i.e.,
\begin{eqnarray}
&&
f^{-1/2}\left(f^{1/2} F_{{\rm S}(\la)}\right)'
\nn
&&
~~~=\,\frac{1}{ r}F_{{\rm S}(\mu)}+
\left(\frac{f'}{ 2 f}-\delta'\right)
F_{{\rm S}(\nu )}+
\phi' F_{{\rm S}(\phi)}
\,.
~~~~~
\end{eqnarray}
That is, the above four equations (\ref{eq_phi}) \,--\, (\ref{eq_mu})
are not independent.
Hence if we solve three of them, the remaining one equation
is automatically satisfied.
We solve the following three equations:
\begin{align}
&
F_{{\rm S}(\phi)}=0\,,
\nn
&
F_{{\rm S}(\nu)}=0\,,
\nn
&
F_{{\rm S}(\nu)}-F_{{\rm S}(\la)}=0\,.
\end{align}
The last equation is explicitly given by
\begin{widetext}
\begin{align}
&r^2 \bigl[(D-2) (h -f'r)
 +4 f (\phi'' -\phi'^2 ) r^2
 -2 (h -f'r - 2 f \phi' r ) \phi' r)\bigr] \nn
&+ \ta_2
 \bigl[
2 (D-4) (k-f) ( h -f' r )
+8 f (k-f) ( \phi'' -2 \phi'^2 ) r^2
 -4 (k-3 f) (h-f' r)\phi'r
\bigr] =0
\,,\label{eq_nu-la}
\end{align}
\end{widetext}
which is simpler than the others.
\section{AdS spacetime}\label{sec3}
First we look for an AdS spacetime for $k=0$.
The metric form is assumed to be
\begin{align}
&f= \frac{r^2}{ \ell^2} \,,~~\delta=0\,,
\label{ansatz_k0}
\end{align}
where $\ell$ is an AdS radius, which will be fixed later.

Assuming the dilaton field as
\begin{align}
\phi=\frac{p}{2}\ln\left(\frac{r^2}{ \ell^2}\right)
\,,
\label{ansatz_k1}
\end{align}
we find the following three algebraic equations:
\begin{align}
&
\left[D_1-4(D-1) p+4 p^2 \right] \ell^2
-\ta_2 D_1=0 ,
\label{3-3}\\
&
\left[
(D-1)_2-4(D-2) p+4 p^2\right] \ell^2
\label{3-4}\\
&-\ta_2
\left[(D-1)(D-4) -8(D-2)p +16 p^2 \right]=0 ,
\nn
&\ell^2-2(1+2p)\ta_2 =0
\label{3-5} \,.
\end{align}
These equations are not independent since \eq{3-5} can be
reproduced with \eq{3-3} and \eq{3-4}.
If we find a solution of these equations
for two unfixed parameters ($\ell$ and $p$),
then we obtain an AdS spacetime.
Eliminating $\ell$ from Eqs. (\ref{3-3}) and (\ref{3-5}),
we find the equation for~$p$:
\begin{eqnarray}
16 p^3-8(2D-3)p^2 +4(D-1)_2 p+D_1=0
\label{eq_cubic}
\,.
\end{eqnarray}
The AdS radius is given by its solution $p$ as
\begin{eqnarray}\label{AdSradius}
\ell=\sqrt{2(1+2p)}\, \ta_2^{1/2}\,.
\end{eqnarray}
We find one real solution for Eq. (\ref{eq_cubic}),
which gives an AdS space for $\a_2>0$.\footnote{For $\a_2<0$,
we find a de Sitter space.}
We show the explicit values for an AdS space in Table \ref{table_1}.
We also find
$\ell\rightarrow \ta_2^{1/2}\approx D\a_2^{1/2}$
and $p\rightarrow -1/4$ as $D\rightarrow \infty$.

\begin{table}
\caption{The AdS radius $\ell$ and the power $p$ of the dilaton field distribution
for $D=4\,$--$\,10$. The power exponent $z=(D-2-2p)/(D-2)$ appears in the Einstein frame.}
\begin{center}
\begin{tabular}{|c|c|c||c|}
\hline
~~$D$~~&~~$\ell/\ta_2^{1/2}$~~&~~$ p$~~&~~$ z$~~
\\
\hline
~~$4$~~&~~$0.86148580$~~&~~$-0.31446055$~~&~~$ 1.31446055$
\\
\hline
~~$5$~~&~~$0.89109384$~~&~~$-0.30148794$~~&~~$ 1.20099196$
\\
\hline
~~$6$~~&~~$0.91035546$~~&~~$-0.29281323$~~&~~$ 1.14640662$
\\
\hline
~~$7$~~&~~$0.92386678$~~&~~$-0.28661754$~~&~~$ 1.11464702$
\\
\hline
~~$8$~~&~~$0.93385873$~~&~~$-0.28197697$~~&~~$ 1.09399232$
\\
\hline
~~$9$~~&~~$0.94154323$~~&~~$-0.27837409$~~&~~$ 1.07953545$
\\
\hline
~~$10$~~&~~$0.94763439$~~&~~$-0.27549727$~~&~~$ 1.06887432$
\\
\hline
\end{tabular}
\label{table_1}
\end{center}
\end{table}

The solution is explicitly written in the line element form:
\begin{align}
ds^2 &= -\frac{r^2}{\ell^2} dt^2
 + \frac{\ell^2}{r^2} dr^2 + r^2 d~\vect{x}_{D-2}^2 ~.
~~ \label{AdS_p} \\
e^\phi &= \left(\frac{r}{\ell} \right)^p,
~~ \label{AdS_phi}
\end{align}
where $d~\vect x_{D-2}^2$ denotes the $(D-2)$-dimensional flat Euclidean space.
It is conformally flat:
\begin{eqnarray}
ds^2&=&\frac{\ell^2}{ u^2}\left[
- d \tau^2+d u^2+d~\vect x_{D-2}^2
\right],
\\
e^\phi &=& u^{-p},
\end{eqnarray}
where we have introduced the dimensionless
variables as $\tau \equiv t/\ell$ and
$u\equiv \ell/r $.

In the next two subsections, we will show
two different descriptions of the present solution.

This type of exact AdS solution is also found in the generalized Lovelock
gravity theory, which is given in Appendix \ref{AdS_Lovelock}.

\subsection{The solution described in the Einstein frame}

To describe our solution in the Einstein frame,
we perform a conformal transformation
\begin{eqnarray}
g_{\E \mu\nu}&=&\exp[ -2\gamma^2 \phi\,]\, g_{\mu\nu},
\label{conf_trans2}
\end{eqnarray}
where $\c=\sqrt{2/(D-2)}$.
Here and in what follows, we use the subscript $E$ for the variables
in the Einstein frame.
Note that the conformal transformation changes  the circumference
radius due to the nontrivial linear dilaton field, i.e.,
\begin{align}
r_\E&=\exp [-\gamma^2 \phi ] \, r , \label{radial2} \\
\Leftrightarrow ~~ \frac{r}{\ell}&= \lh \frac{r_\E}{\ell}
 \rh^{\frac{1}{z}} ,\qq
z \equiv \frac{D-2-2p}{D-2}.
\label{psize}
\end{align}
The value $z$ is given in Table \ref{table_1}.
Hence the line element in the Einstein frame becomes
\begin{align}
ds_\E^2 = -\lh \frac{r_\E}{\ell} \rh^2 dt^2
 +z^{-2} \lh \frac{\ell }{r_\E} \rh^{\frac{2}{z}} dr_\E^2
 + r_\E^2 d~\vect x_{D-2}^2
\end{align}

By introducing new variables as
$\tau_\E \equiv t/\ell$ and $ u_\E\equiv \left( \ell/r_E \right)^{1/z}$,
we find
\begin{align}
ds_\E^2=\frac{\ell^2}{ u_\E^{2 z}}\left[
- d \tau_\E^2+d u_\E^2+d~\vect x_{D-2}^2
\right] .
\end{align}
It is conformally flat but no longer an AdS spacetime.
The dilaton field is also nontrivial, i.e.,
\begin{align}
e^\phi= u_\E^{-p}
\,.
\end{align}

\subsection{Hyperbolic chart of AdS spacetime}

Next we look at the other chart of the AdS spacetime.
There are three static charts of AdS spacetime depending on their
spatial curvatures ($k=0, \pm 1$). The previous analytic solution
corresponds to a flat slicing ($k=0$).
We can change the chart from this plane symmetric one
to others by a coordinate transformation.
Here we present one example, which is the chart of the hyperbolic slicing.
Consider the transformation
\begin{align}
t&= \bar r \lh 1 - \frac{\bar r^2 }{ \ell^2} \rh^{-1/2}
e^{\bar t / \ell}~\cosh{\theta}~, \\
r&=\ell \lh 1 - \frac{\bar r^2 }{ \ell^2} \rh^{1/2} e^{-\bar t / \ell}~,\\
x_{j} &= \frac{\bar r}{\ell} \lh 1 - \frac{\bar r^2 }{ \ell^2} \rh^{-1/2}
e^{\bar t / \ell}~\sinh{\theta}~ \\
&\times\lh\prod_{k=1}^{j} \sin{\phi_{k-1}}\rh \cos{\phi_{j}},
 \quad \lh ~j=1,\cdots, D-2 \rh \nonumber
\end{align}
where we impose $\sin{\phi_0}=\cos{\phi_{D-2}}=1$.
This gives
\begin{align}
ds^2 =& -\lh -1 + \frac{\bar r^2}{\ell^2} \rh d \bar t^2 \nn
 & + \lh -1 + \frac{\bar r^2}{\ell^2} \rh^{-1} d \bar r^2
 + \bar r^2 d\Sigma_{-1}^2 \label{AdS_h} ~,
\end{align}
where $d\Sigma_{-1}^2$ denotes the line element
of $(D-2)$-dimensional hyperbolic space.
AdS spacetime in this chart is often referred to as a massless
AdS black hole with the horizon radius $\bar r_H= \ell$
~\cite{Birmingham}, although it is the AdS horizon.

Although this spacetime looks static, in this coordinate system,
the dilaton field is not static but time-dependent
i.e.,
\begin{align}
\phi = \frac{p}{2} \ln
\left( 1 - \frac{\bar r^2 }{ \ell^2} \right)
- p \frac{\bar t}{ \ell } ~.
\end{align}
It is because the dilaton field is static but inhomogeneous in the flat
slicing, and then it becomes time-dependent as well as inhomogeneous after
changing the time slicing to the hyperbolic one.

Spherical case is easily obtained by the replacement
$\bar r \to i \hat r$, $\bar t \to i\hat t$, and $\bar \theta \to i\hat \theta$:
\begin{align}
ds^2 =& -\lh 1 + \frac{\hat r^2}{\ell^2} \rh d \hat t^2
 + \lh 1 + \frac{\hat r^2}{\ell^2} \rh^{-1} d \hat r^2
 + \hat r^2 d\Sigma_{1}^2 \label{AdS_s} ~, \\
\phi =& \frac{p}{2} \ln
\left( 1 + \frac{\hat r^2 }{ \ell^2} \right)
- ip \frac{\hat t}{ \ell } ~,
\end{align}
though the dilaton is not real. 
\section{Boundary conditions}

Before we solve Eqs. \eq{eq_phi}, \eq{eq_nu} and \eq{eq_nu-la}
to find a black hole solution, we need the boundary conditions
at an event horizon and at an infinity.
We discuss those separately.
\subsection{Boundary conditions on the horizon}
At the event horizon ($r_H$), the metric function $f$ vanishes,
i.e., $ f(r_H)= 0$. The metric functions, the dilaton field, and their derivatives
must be finite at $ r_H$. Since $X$ has $f$ in the denominator,
we find nontrivial relations between the variables at the horizon
as discussed in \cite{ohta_torii1,maeda_ohta_sasagawa}.
Expanding the basic equations near the horizon
and taking the limit of $r\rightarrow r_H$,
we obtain the following
three independent constraints from
regularity conditions on the horizon:
\begin{align}
& \rho_H^2
\left[(D-2)_{3}k -2(D-2)\xi_H+2 \zeta_H+4\xi_H \eta_H\right] \nn
&~+\left[(D-4)_{5}k^2 -4(D-4)k\xi_H+4 k\zeta_H+2\xi_H^2\right]=0,
\label{BC1}
\\
& \rho_H^2
\left[(D-2)_{3}k-(D-2)\xi_H+2\xi_H \eta_H \right] \nn
&~+\left[(D-4)_{5}k^2 -2(D-4)k\xi_H+4 k\xi_H \eta_H\right]=0,
\label{BC2}
\\
&(D-2) \rho_H^2
\left[(D-3)_{4}k-2(D-3)\xi_H+2 \zeta_H+4\xi_H \eta_H \right]
\nn
&~+\Bigl[(D-4)_{6}k^2 -4(D-4)_{5}k\xi_H+4(D-4)k\zeta_H \Bigr. \nn
&~~\Bigl. +8(D-4)k\xi_H \eta_H+2(D-4)k\xi_H^2-8\xi^2 \eta_H \Bigr]=0,
\label{BC3}
\end{align}
where we have denoted the variables at the horizon with the subscript $H$,
i.e., $ \phi_H$, $ \phi_H'$,
$ f_H'$,$ \delta_H$,
$ \delta_H'$,
$ ( X f)_H$, and so on,
and introduced the dimensionless variables as
\begin{align}
 \rho_H&:= r_H/\ta_2^{1/2}~,~~&
\xi_H&:= r_H f_H'~,~~ \nn
 \eta_H&:= r_H \phi_H'~,~~ &
 \zeta_H&:= r_H^2 ( X f)_H
\,.
\end{align}
Eliminating $\eta_H$ and $ \zeta_H$ in Eqs.
(\ref{BC1}) \,--\, (\ref{BC3})
(we assume that $\xi_H\neq 0$ and $ \zeta_H\neq 0$),
we find the quadratic equation for $ \xi_H$:
\begin{eqnarray}
 a \xi_H^2 + b \xi_H+ c=0
\label{eq_quad}
\,,
\end{eqnarray}
where
\begin{align}
 a=&2 ( \rho_H^2 +2 k ) ( (D-1) \rho_H^2 + 2 (D-4) k ) ,
\label{reg_quad} \\
 b=&- \bigl[ (D-2) \rho_H^6 +4 ( D^2 - 3 D -1 ) k\rho_H^4 \bigr. \\
&~~~\bigl. +2 (D-4) (7D -9) k^2\rho_H^2 +4 (D-4) (3D-11) k^3 \bigr],
\nn
 c=& k \bigl[ 4(D-3)_5 k^3 +2(D-4) (3D^2-14D+7) k^2 \rho_H^2 \bigr. \nn
 &\bigl. +2(D^3-5D^2+2D+14) k\rho_H^4 +(D-2)_3 \rho_H^6 \bigr].
\end{align}

As discussed in \cite{maeda_ohta_sasagawa},
the discriminant of the quadratic equation~(\ref{eq_quad}), which depends
on $D$ and $\rho_H$, must be non-negative for the existence of a regular horizon.
Fixing a fundamental coupling constant $\ta_2$ gives
constraints on the horizon radius $ r_H$ for given $D$.

Assuming that a regular horizon exists,
we find two branches: the plus and minus branches
\begin{align}
\xi_{H\pm} = \frac{-b \pm \sqrt{b^2-4 ac}}{2 a}.
\end{align}
In the following, we discuss three cases separately by the horizon topology
(or $k=0,\pm1$).

\subsubsection{Planar symmetric case {\rm ($k=0$)}}
The quadratic equation \eq{eq_quad} is reduced to
\begin{align}\label{eq_quad_k0}
\left[ 2(D-1)\xi_H - (D-2)\rho_H^2 \right] \xi_H =0
\,.
\end{align}
There exists only one positive real solution, which is
\begin{align}
\xi_H=\frac{(D-2)}{(D-1)} \frac{\rho_H^2}{2} .
\end{align}
Although $\xi_H=0$ is also the solution of \eq{eq_quad_k0},
it does not give a proper horizon because the expansion around the
horizon reveals that all derivatives of
$f$ vanish in this case.

\subsubsection{Spherically symmetric case {\rm ($k=1$)}}

Assuming $\ta_2>0$ in $D=4$\,--\,$10$,
the allowed values for the regular event horizon
are summarized in Table II in \cite{maeda_ohta_sasagawa}
and the real $\xi_H$ is always positive.
In the plus branch, we obtained asymptotically flat black hole solutions
for some range of parameters given in Table VI
in \cite{maeda_ohta_sasagawa},
while we have found a curvature singularity in the other cases
including the negative branch.
Here we also find a solitonic solution with a regular center and no
 horizon in one of the branches (see Appendix \ref{solitonic}).

\subsubsection{Hyperbolically symmetric case {\rm ($k=-1$)}}

Assuming $\ta_2>0$ in $D=4$\,--\,$10$,
we find the allowed values for a regular event
horizon, which are summarized in Table \ref{table_2}.
In any dimensions higher than three, we always find a small
gap in the parameter space of horizon radius,
where there is no regular horizon.
It makes a clear difference from the case of $k=1$;
there is a minimum horizon radius in four dimensions,
a small gap in five and six dimensions, and
no restriction in dimensions higher than six~\cite{maeda_ohta_sasagawa}.
We find some ranges where $\xi_H$ become negative for $k=-1$.
Since a horizon with negative $\xi_H$ no longer describes
a black hole horizon but a cosmological horizon,
further restrictions arise for the range of a regular black hole horizon.
The solution can be interpreted as a cosmological one.
We present this cosmological solution in Appendix \ref{cosmological}.

It may be worth introducing an important horizon radius
\begin{align}
r_H^2=\frac{2(D-4)}{(D-1)} \ta_2,
\label{upper_bound}
\end{align}
at which the coefficient of the quadratic term in \eq{eq_quad} vanishes.
It is the bifurcation point where the sign of $\xi_H$ changes.
This gives the upper bound of the range for which the asymptotically AdS black
hole solutions can be obtained numerically, as we show in the next section.

\begin{table}[h]
\caption{The allowed range of horizon radii for a regular horizon in the case of $k=-1$.
In any dimension higher than three, there is a gap
where there is no regular horizon.
}
\begin{center}
\begin{tabular}{|c|c|}
\hline
$D$& ~~Condition for regular horizon with $k=-1$
\\
\hline
4&~~$ r_H\leq 1.6566722 \, \ta_2^{1/2} \,~~
{\rm or}~~~ 2.4494897 \, \ta_2^{1/2} \leq r_H$~~
\\
\hline
5&~~$ r_H\leq 1.5458773 \, \ta_2^{1/2}\,~~
{\rm or}~~~ 2.6272504 \, \ta_2^{1/2}\leq r_H$~~
\\
\hline
6&~~$ r_H\leq 1.4951145 \, \ta_2^{1/2}\,~~
{\rm or}~~~ 2.7522118 \, \ta_2^{1/2}\leq r_H$~~
\\
\hline
7&~~$ r_H\leq 1.4684167 \, \ta_2^{1/2}\,~~
{\rm or}~~~ 2.8514273 \, \ta_2^{1/2}\leq r_H$~~
\\
\hline
8&~~$ r_H\leq 1.4528824 \, \ta_2^{1/2}\,~~
{\rm or}~~~ 2.9349158 \, \ta_2^{1/2}\leq r_H$~~
\\
\hline
9&~~$ r_H\leq 1.4431206 \, \ta_2^{1/2}\,~~
{\rm or}~~~ 3.0076296 \, \ta_2^{1/2}\leq r_H$~~
\\
\hline
10&~~$ r_H\leq 1.4366113 \, \ta_2^{1/2}\,~~
{\rm or}~~~ 3.0724201 \, \ta_2^{1/2}\leq r_H$~~
\\
\hline
\end{tabular}
\label{table_2}
\end{center}
\end{table}

\subsection{Boundary condition at spatial infinity
and ``mass" term
 } \label{boundary}
In order to impose the boundary condition at infinity,
we assume that spacetime approaches the solution
(\ref{ansatz_k0}) with (\ref{ansatz_k1}) as
$r\rightarrow \infty$.

Since we look for black hole solutions, we may have to find the mass.
For the Schwarzschild AdS black hole, the metric is given by
$\delta=0$ and
\begin{eqnarray}
f(r)=r^2/\ell^2+k+\mu/r^{D-3}
\,,
\end{eqnarray}
where $\mu$ is the conserved Arnowitt-Deser-Misner mass.
So we expect the mass term may appear in a negative power of $r/\ell$.

It may be natural to assume
analyticity of spacetime at infinity, which guarantees
that the spacetime is really asymptotically flat or AdS.
Hence, assuming analyticity of spacetime at infinity,
we first expand the variables around the AdS spacetime
by an integer-power series of $r/\ell$ as
\begin{eqnarray}\label{expansion}
f (r) &=& \bar f(r):=\left( \frac{r}{\ell} \right)^{2}
\left[1+\sum_{n=1}^\infty
f_n \left( \frac{r}{ \ell} \right)^{-n}\right]
\,,
\label{expansion_f}
\\
e^{\d (r)} &=& e^{\bar\d (r)}:= e^{\d_0}\left[1 + \sum_{n=1}^\infty
\d_n \left( \frac{r}{ \ell} \right)^{-n}\right]
\,,
\label{expansion_d}
\\
e^{\phi (r)} &=&e^{\bar \phi (r)} :=
 \left( \frac{r}{ \ell} \right)^{p}
\left[1 + \sum_{n=1}^\infty
\phi_n \left( \frac{r}{ \ell} \right)^{-n}\right].
\label{expansion_phi}
\end{eqnarray}
With the use of a gauge freedom of time coordinate,
we can set $\d_0=0$.

Inserting the expression (\ref{expansion}) into the basic equations,
we find the equations for the first-order perturbations as
\begin{align}
&(D-2-2p) \bigl[\bigr. p (D-3-2p) f_1 \label{perturb2_phi4} \nn
&\hspace{20mm} +2p \d_1 +2(1+2p)  \phi_1 \bigl. \bigr]
 =0 \\
&(D-2-2p) \bigl[ (1+Dp-2p^2)f_1
 +4 p \phi_1 \bigr]
 =0 \label{perturb2_nu4} \\
& \bigl[ p (1 + 2 p) f_1
 + (1+Dp-2p^2)  \d_1
\bigr]
 =0\,,
 \label{perturb2_nu_la4}
\end{align}
where we have used a relation \eq{AdSradius}.
This gives a trivial solution, i.e.,
 $f_1=\d_1=\phi_1=0$.

Hence we go further to the second-order terms, which should satisfy
\begin{align}
&(D-2)(D-3) p k - p (D-3-2p)(D-4-2p) f_2 \nn
&~~~-4 p (D-3-2p) \d_2 \nn
&~~~-4 (1+2p) (D-3-2p) \phi_2
 =0 , \label{perturb2_phi1} \\
& [ (D-3)(1+Dp) -2p(1+2p) ]k \nn
&~~~-(D-3-2p)(1+Dp-2p^2) f_2 \nn
&~~~ -8 p (D-3-2p) \phi_2
 =0 \ , \label{perturb2_nu1} \\
& p(1 + 2 p) k + p (1 + 2 p) f_2 \nn
&~~~- 2 (1+Dp-2p^2) \d_2 -8 p \phi_2
 =0 \ .
\label{perturb2_nu_la1}
\end{align}
These equations are solved as
\begin{align}
f_2 &= k + 4 N [1+(D+2)p-2(D-4)p^2] \ , \nn
\d_2 &=-2 N (D-3) [1 + (D+2)p + 4 p^2] \ , \nn
\phi_2 &=\frac{k p}{ 2} - N \left[
 1 + 3 D p + 2 (D^2+2) p^2 + 8 p^3 \right] \ ,
\end{align}
with
\begin{align}
N=\frac{k p^2 }{ \left[2(D-3-2p) (1+Dp)(1+(D+2)p+2p^2)
\right]} \,,
\end{align}
whose explicit values are given in Table \ref{table_3}.
\begin{table}[h]
\caption{The coefficients of the second-order
perturbations $f_2, \d_2,$ and $\phi_2$ for $k=-1$ in $D=4$\,--\,$10$.
As $D$ gets higher, $f_2$ approaches the value of the curvature constant $k$.
Note that $f_2=k$ in the EGB model without a dilaton field.
}
\begin{center}
\begin{tabular}{|c|c|c|c|}
\hline
~~$D$~~&~~$f_2$~~&~~$\d_2$~~&~~$\phi_2$~~
\\
\hline
~~$4$~~&~~$-0.39395966$~~&~~$ -0.16785769$~~&~~$ 0.065379959$~~
\\
\hline
~~$5$~~&~~$-0.80848349$~~&~~$ -0.11068778$~~&~~$ 0.10751215$~~
\\
\hline
~~$6$~~&~~$-0.90905634$~~&~~$ -0.080899823$~~&~~$ 0.11882449$~~
\\
\hline
~~$7$~~&~~$-0.94772138$~~&~~$ -0.063112022$~~&~~$ 0.12330545$~~
\\
\hline
~~$8$~~&~~$-0.96634517$~~&~~$ -0.051448744$~~&~~$ 0.12540494$~~
\\
\hline
~~$9$~~&~~$-0.97664170$~~&~~$ -0.043278079$~~&~~$ 0.12647678$~~
\\
\hline
~~$10$~~&~~$-0.98289482$~~&~~$ -0.037266878$~~&~~$ 0.12704557$~~
\\
\hline
\end{tabular}
\label{table_3}
\end{center}
\end{table}

Even when we go beyond the second order, we find that
nontrivial even power terms are given only by known parameters
($D, k$, and $p$).
If one regards the coefficient of the term proportional to $r^{-(D-3)}$
as a mass, we find a finite mass for odd dimensions
(see Table \ref{table_7}) but zero mass for even dimensions.
\begin{table}
\caption{The coefficient of the term proportional to $r^{-(D-3)}$ for $k=-1$.}
\begin{center}
\begin{tabular}{|c|c|c|}
\hline
$D$& ~~~$ f_{D-1} $~~~ & ~~~$\phi_{D-1}$~~~
\\
\hline
5&~~~~$0.44468899$~~~~&~~~~$-0.0015225937$~~~~
\\
\hline
7&~~~~$0.064294258$~~~~&~~~~$0.039985499$~~~~
\\
\hline
9&~~~~$0.025679692$~~~~&~~~~$0.048778008$~~~~
\\
\hline
\end{tabular}
\label{table_7}
\end{center}
\end{table}
Moreover, the values are fixed only by
the fundamental constants of the theory ($D$ and $\ta_2$).
There is no additional free parameter such as the conventional mass
or conserved charge,
which depends on the horizon radius.
Hence we conclude that under the present expansion ansatz,
important quantities of a black hole such as the mass are not involved.
Note that this expansion gives the exact AdS spacetime
in the plane symmetric case ($k=0$), while for $k\neq 0$, we obtain nontrivial
deviation from the AdS spacetime. However, it turns out that
any solution with this asymptotic expansion terms is no longer regular.
We need extra asymptotic terms to obtain a regular solution
 (see Appendix \ref{solitonic}).

In order to extract a kind of mass parameter
or other properties of a black hole,
which depends on a horizon radius,
we have to assume nonanalytic expansion at infinity as
\begin{align}
f(r)=\bar f(r)-{\mu_f}\left({r\over \ell}\right)^{-\nu_{f}}+\cdots \,, \\
\phi(r) = \bar \phi(r) +\mu_{\phi}
\left({r\over \ell}\right)^{-\nu_\phi}+\cdots \,, \\
\d(r) = \bar \d(r) +{\mu_{\d}}\left({r\over \ell}\right)^{-\nu_\d}
+\cdots \,.
\end{align}
where $\bar f(r)$, $\bar \phi(r)$ and $\bar \d(r)$ are given by
Eqs.~\eq{expansion_f},
\eq{expansion_phi} and \eq{expansion_d}, respectively,
and $\nu_f$, $\nu_{\phi}$, and $\nu_{\d}$ are positive but noninteger numbers.
Inserting this expression into the basic equations,
we find
\begin{align}
\nu_{f}=D-3-2p\,,~~\nu_{\phi}=\nu_{\d}=D-1-2p.
\end{align}
$\mu_f$, $\mu_{\phi}$ and $\mu_{\d}$ are constants,
 which satisfy one constraint condition:
\begin{align}
p(1+2p) &\mu_f +(D-1-2p) (1+Dp-2p^2)\mu_{\d} \nn
 &+2p(D-1-2p)(D-2-2p) \mu_{\phi} =0
\,. \label{perturb_const}
\end{align}
Hence two of them are free and represent some properties of a black hole.
We call $\mu_f$ and $\mu_{\phi}$ a ``mass" and a ``scalar charge",
respectively, although they may not be a proper mass and
a proper scalar charge.
In \cite{ohta_torii2}, there exists a non-integer power
``mass" term similar to our case,
though it was observed that the correct integer power of the mass term is restored
by taking into account the effect of lapse function $\delta$.
The AdS spacetime in our system is realized by the presence of a nontrivial
dilaton field and is completely different from \cite{ohta_torii2},
where a negative cosmological constant is introduced.
%
%

\section{Asymptotically AdS black hole solutions}
\subsection{Numerical solution}

Now we present asymptotically AdS black hole solutions.
Giving a horizon radius $r_H$, we solve the basic equations
(\ref{eq_phi}), (\ref{eq_nu}) and (\ref{eq_nu-la}) numerically.
To integrate the equations,
we first set $\delta_H=0$ and $\phi_H=0$ and find
the asymptotically AdS spacetime with a linear dilaton field.
We then rescale the lapse function and the dilaton
field as $\tilde \d (r)=\d (r)-\d (\infty)$
and $\tilde \phi(r)= \phi(r) -
 [ \phi(r) - p/2 \ln \left( r^2/\ell^2 \right) ] |_{r=\infty}$, respectively.
This is always possible because $\delta$ and
$\phi$ appear only in the forms of their
derivatives such as $\delta'$ and $\phi'$.
As a result, we find $\tilde \d (r) \rightarrow 0$ and 
$\tilde \phi(r) \rightarrow p/2 \ln \left( r^2/\ell^2 \right) $
as $r\rightarrow \infty$.
Then in our physical solution with a tilde, $\tilde \delta_H$ and
$\tilde \phi_H$ do not vanish.
In what follows, for brevity, we omit the tilde of the variables.

For $k=0$, there is no asymptotically AdS black hole solution.
The reason is as follows:
For a flat 3-space,
there exists an additional rescaling symmetry under
\begin{eqnarray}
f(r) &\rightarrow& f^* (r^*)=
\Lambda^2 f(r),
\nonumber \\
\phi(r) &\rightarrow& \phi^* (r^*)=
\phi(r),
\nonumber \\
h(r) &\rightarrow& h^* (r^*)=\Lambda^2 h(r),
\nonumber \\
 X(r) &\rightarrow& X^* (r^*)=\Lambda^{-2} X(r),
\end{eqnarray}
as $r \rightarrow r^* =\Lambda r$, where $\Lambda$ is an arbitrary constant.
Hence, once we obtain one black hole solution, we can construct one parameter
family of solutions with arbitrary horizon radii by the above rescaling.
However, numerical analysis adopting the boundary condition
with $\rho_H =1$ does not provide a regular solution but encounters a naked singularity.
Therefore, we conclude that there exists no regular black hole solution for $k=0$.
The only regular solution is the exact AdS spacetime with the linear dilaton field.
For $k=1$, there is no asymptotically AdS black hole solution:
In \cite{maeda_ohta_sasagawa}, we solved the field equations outward from horizon
without imposing a boundary condition at infinity
and found that all the regular solutions are asymptotically flat.
This means that there is no asymptotically AdS black hole solutions for $k=1$
at least within the numerical analysis.
However, we find a solitonic solution, which is regular everywhere including the origin
(see Appendix \ref{solitonic}).

Hereafter we focus on the case of $k=-1$.
For $D>4$, we find an asymptotically AdS black hole solution.
Depending on the dimension $D$, we classify our solutions into two types:
type (a), five dimensions ($D=5$), in which we find an upper bound
in the range of horizon radius but no lower bound; and
type (b), six or higher dimensions ($D=6$\,--\,$10$),
for which there is a finite range of the horizon radius.

\begin{figure}[b]
\begin{minipage}{50mm}
\begin{center}
\includegraphics[width=50mm]{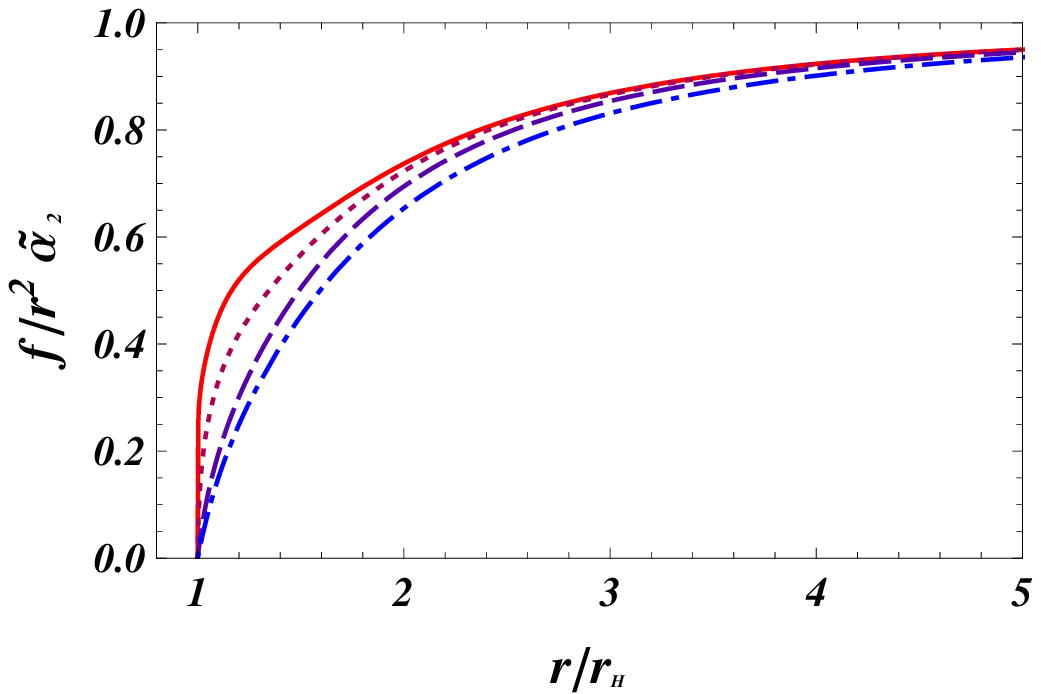} \\
(a) metric function $f$ \\
\end{center}
\vspace{-4mm}
\begin{center}
\includegraphics[width=50mm]{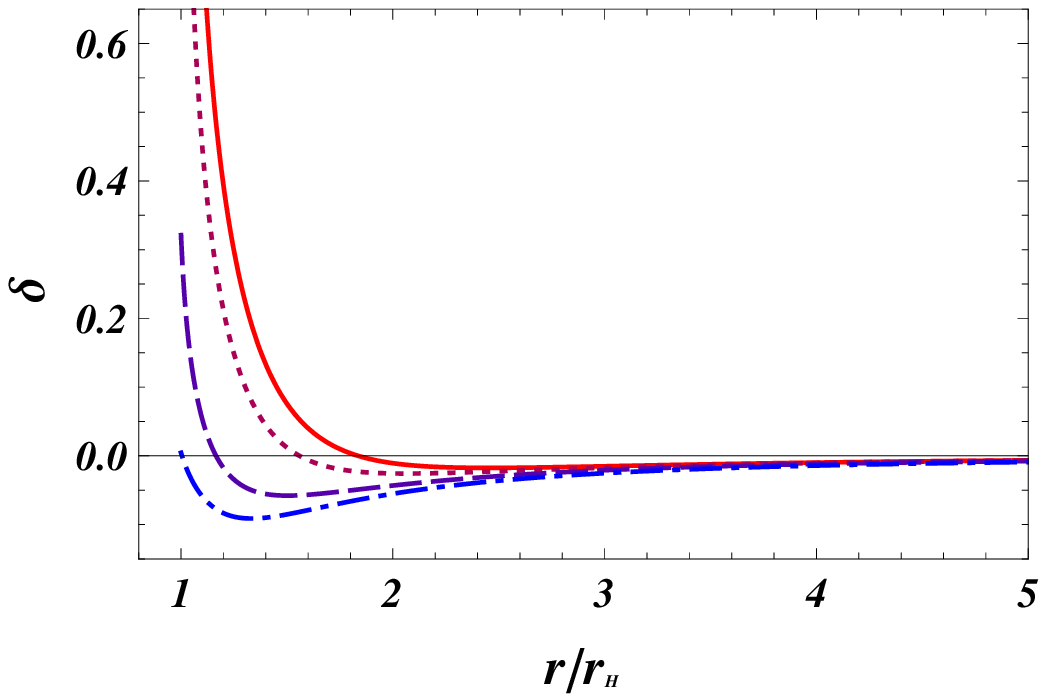} \\
(b) lapse function $\delta$\\
\end{center}
\vspace{-4mm}
\begin{center}
\includegraphics[width=50mm]{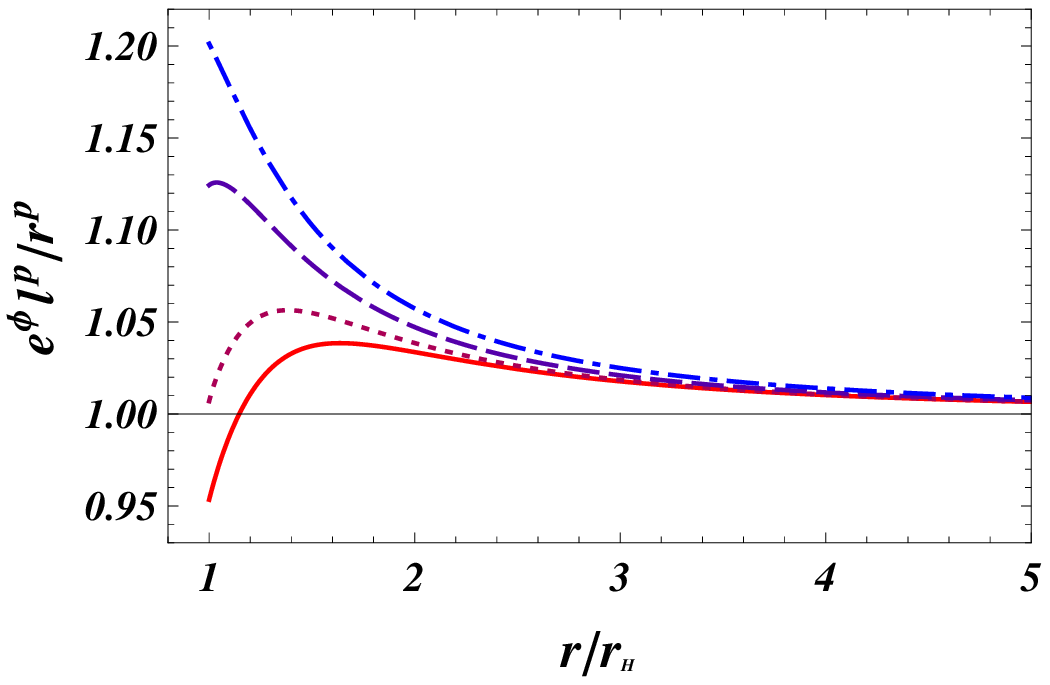}\\
(c) dilaton field $\phi$
\end{center}
\end{minipage}
\caption{The metric functions $f$ and $\delta$ and the
dilaton field $\phi$ in five dimensions for $k=-1$.
We plot them normalized by the AdS values.
We choose the following four values for the horizon radius:
$r_H=$\, 0.70704304$\,\ta_2^{1/2}$ (solid line),
 0.70353475$\,\ta_2^{1/2}$ (dotted line),
0.67226423$\,\ta_2^{1/2}$ (dashed line),
and 0.62013066$\,\ta_2^{1/2}$ (dot-dashed line).
}
\label{sol5}
\end{figure}

\begin{figure}[h]
\begin{minipage}{50mm}
\begin{center}
\includegraphics[width=50mm]{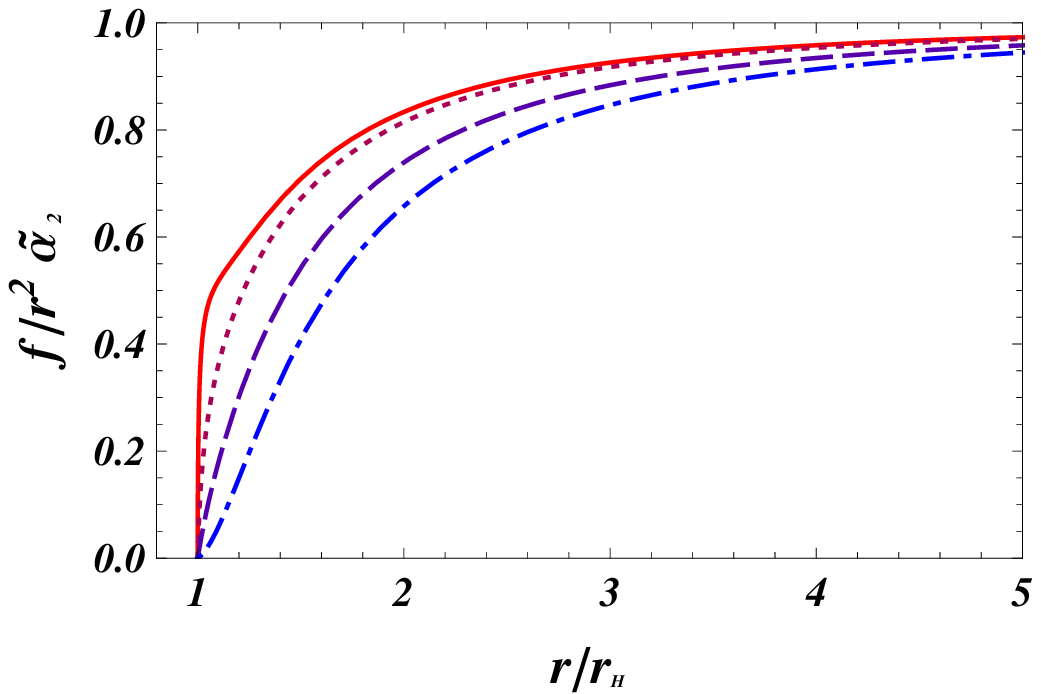} \\
(a) metric function $f$ \\
\end{center}
\vspace{-4mm}
\begin{center}
\includegraphics[width=50mm]{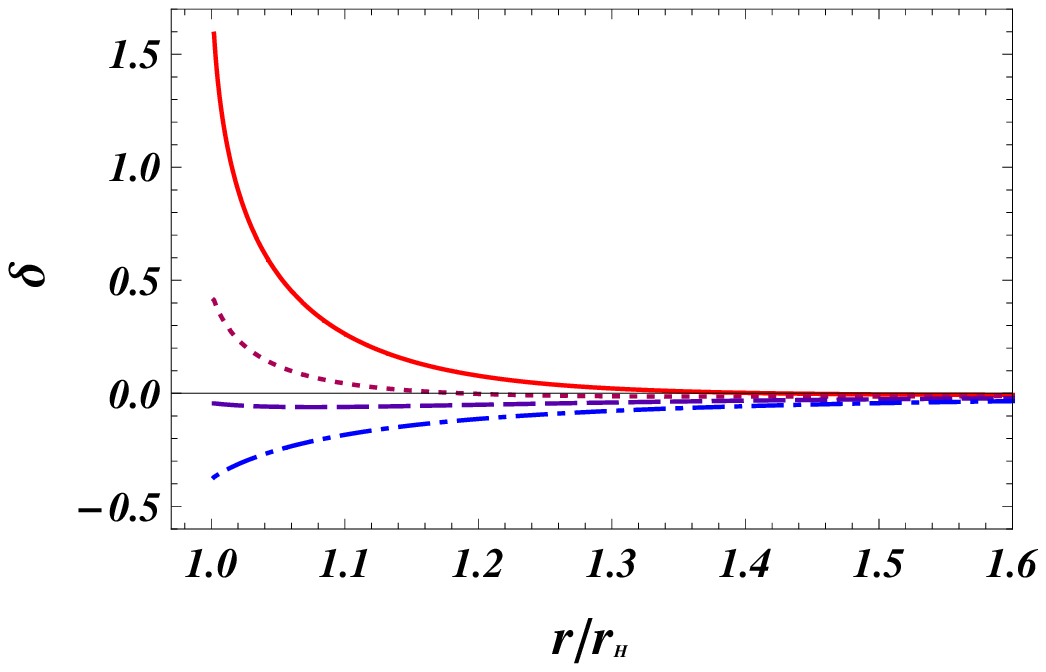} \\
(b) lapse function $\delta$\\
\end{center}
\vspace{-4mm}
\begin{center}
\includegraphics[width=50mm]{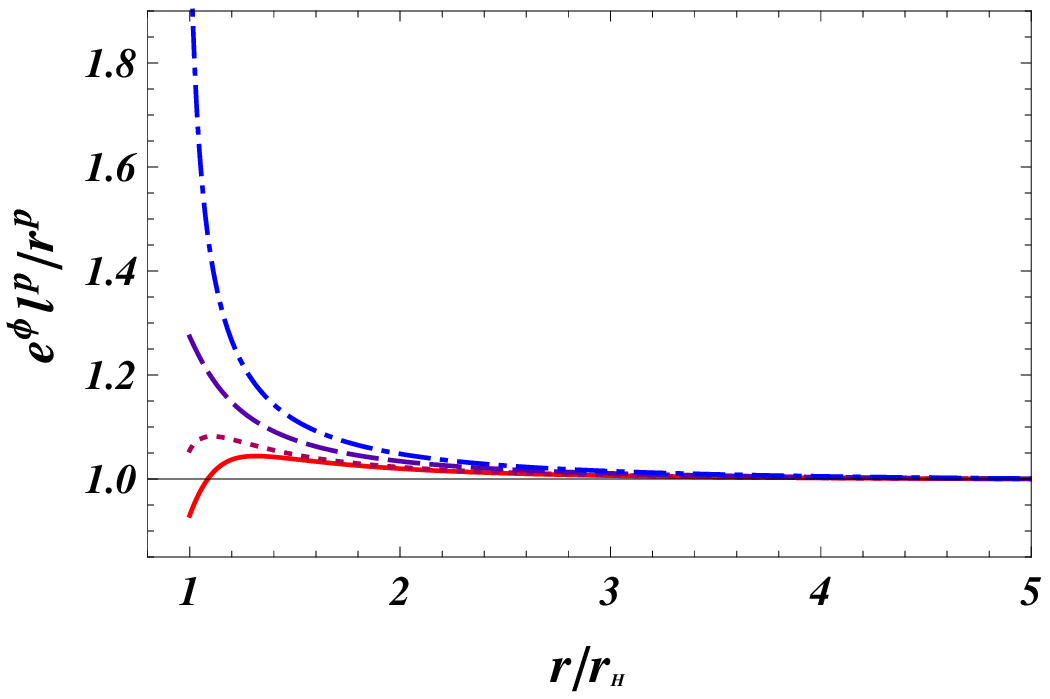}\\
(c) dilaton field $\phi$
\end{center}
\end{minipage}
\caption{The metric functions $f$ and $\delta$ and the
dilaton field $\phi$ in ten dimensions for $k=-1$.
We plot them normalized by the AdS values.
We choose the following four values for the horizon radius:
$r_H=$\, 1.1486586$\,\ta_2^{1/2}$ (solid line),
1.0894228$\,\ta_2^{1/2}$ (dotted line),
0.9164287$\,\ta_2^{1/2}$ (dashed line),
 and 0.7976378$\,\ta_2^{1/2}$ (dot-dashed line).
}
\label{sol10}
\end{figure}

In Figs~\ref{sol5} and \ref{sol10},
we depict solutions for $D=5$ and 10.
The metric function $f$ and the dilaton field $\phi$
are normalized by the AdS values,
which is given by Eqs. (\ref{ansatz_k0}) and (\ref{ansatz_k1}).
The metric function monotonically approaches the asymptotic value
in both five and ten dimensions.
In five dimensions the lapse function decreases near the horizon
and increasingly gets close to the asymptotic value, while in ten dimensions
it monotonically gets close to the asymptotic value.


The range for the horizon radius of AdS black hole solutions
is summarized in Table \ref{table_4}.
No regular asymptotically AdS black hole solution is found in four dimensions.

In five dimensions, there is no lower bound but exists the upper
bound, i.e., $ 0 < r_H \leq 0.707087\, \ta_2^{1/2}$.
Note that there is no regular solution without a horizon ($r_H=0$).
In dimensions higher than five, we find both lower and upper bounds.
For example, in six and ten dimensions, we find $ 0.514410\,
\ta_2^{1/2} \leq r_H \leq 0.894427\,\ta_2^{1/2}$ and $ 0.782926\, \ta_2^{1/2} \leq
 r_H \leq 1.154671\, \ta_2^{1/2} $, respectively.
The maximum bound on the horizon radius is given by $r_H=\sqrt{2(D-4)/(D-1)}\, \ta_2^{1/2}$
in \eq{upper_bound},
at which the quadratic term in the condition \eq{eq_quad} for a regular horizon vanishes.

Comparing Table~\ref{table_4} with Table~\ref{table_2},
we find that the branches with larger horizon radii in Table~\ref{table_2}
do not give an asymptotically AdS black hole solution, but only
the smaller branches do.

\begin{table}[h]
\caption{The range of the horizon radius in which
asymptotically AdS black holes exist for $k=-1$.
}
\begin{center}
\begin{tabular}{|c|c|}
\hline
$D$& The range of horizon radius~~~
\\
\hline
\hline
4&~~~~$ {\rm None} $~~~~
\\
\hline
5&~~~~$ 0.000000 < r_H/\ta_2^{1/2} \leq 0.707087 $~~~~
\\
\hline
6&~~~~$ 0.514410 \leq r_H/\ta_2^{1/2} \leq 0.894427 $~~~~
\\
\hline
7&~~~~$ 0.636016 \leq r_H/\ta_2^{1/2} \leq 0.999969 $~~~~
\\
\hline
8&~~~~$ 0.705191 \leq r_H/\ta_2^{1/2} \leq 1.069014 $~~~~
\\
\hline
9&~~~~$ 0.750436 \leq r_H/\ta_2^{1/2} \leq 1.118003 $~~~~
\\
\hline
10&~~~~$ 0.782926 \leq r_H/\ta_2^{1/2} \leq 1.154671 $~~~~
\\
\hline
\end{tabular}
\label{table_4}
\end{center}
\end{table}
The results obtained here show a sharp difference from the case of EDEGB
without a cosmological constant, where no AdS black hole solutions
are obtained for the asymptotic constant dilaton~\cite{ohta_torii4}.
The difference comes from the nontrivial dilaton field we consider here.

This solution is a dilatonic analogue of the non-general relativity branch or ``AdS branch" of
the black hole solution in the nondilatonic EGB theory; the
asymptotically AdS black hole solution in the nondilatonic theory~\cite{BH_EGB} is given by
\begin{align}
&
f(r):= k + \frac{(D-2)r^2}{2(D-4)\ta_2} \left( 1
 + \sqrt{1 + \frac{8 (D-4) \ta_2 \mu }{ (D-2)r^{D-1} }} \right), \nn
&\d(r)=0
\,.
\label{AdS_EGB}
\end{align}
The asymptotic behaviour or the weak coupling limit,
i.e., $\ta_2 \mu/r^{D-1} \ll 1$ of this solution~\eq{AdS_EGB}, takes the form
\begin{align}
f(r) &\rightarrow k - \left[\frac{2\kappa_D^2 }{ (D-2)A_{D-2}}\right]
 \frac{M}{ r^{D-3}} + \frac{(D-2) r^2}{(D-4) \ta_2},
\label{asym_EGB}
\end{align}
where we have introduced the area of $N$-dimensional unit sphere $A_N$
and gravitational mass $M$ as
\begin{align}\label{areaN}
A_N&=2\pi^{(N+1)/2}/\Gamma[(N+1)/2], \\
M&= -\frac{ (D-2) A_{D-2}}{\kappa_D^2} \mu.
\label{massEGB}
\end{align}
The asymptotic form \eq{asym_EGB} is the one for an
AdS Schwarzschild solution with a negative mass,
and it properly describes a black hole solution when it satisfies
$f'(r_H)>0$ for $f(r_H)=0$.
This condition is satisfied for $k=-1$ and $r_H$ in the range
\begin{align}
\frac{(D-4)(D-5)}{(D-2)(D-3)} \, \ta_2\leq r_H^2
\leq \frac{2(D-4)}{(D-2)}\,\ta_2
\,.
\label{non_dupper}
\end{align}
The upper bound~\eq{upper_bound} of the horizon radius for the dilatonic solution
is smaller than that in the above inequality~\eq{non_dupper} for nondilatonic case.
In fact, we find that the allowed range of horizon radius of our present black holes
is always included in that in the EGB model, comparing
Table~\ref{table_4} with \eq{non_dupper}.



\subsection{Thermodynamical properties of numerical solution}

Next we present the thermodynamical variables of the AdS black holes.
We can define the temperature of black holes as a period
of Euclidean time coordinate at event horizon
\begin{align}
T_H = \frac{1}{4\pi}f'_H e^{-\d_H},
\end{align}
where $f'_H$ and $\d_H$ are determined by fixing the horizon
radius $r_H$. For the asymptotically AdS solution
in EGB \eq{AdS_EGB}, the temperature is explicitly given by
\begin{align}
T_{H\rm{(EGB)}} = \frac{\left[ (D-2)_3 r_H^2 - (D-4)_5 \ta_2 \right]}
{4 \pi r_H \left[ (D-2) r_H^2 - 2(D-4) \ta_2 \right]}
\,.
\end{align}
Since we consider the higher-curvature correction terms,
we apply Wald's formula for a black hole entropy,
which is generalization of the Bekenstein-Hawking formula
to the case of most generic covariant theory.
Wald's formula is defined by use of the Noether charge
associated with diffeomorphism invariance of a system~\cite{Wald_formula}.
We find
\begin{align}
S = -2 \pi \int_\Sigma \frac{\pa {\cal L}}{\pa R_{\mn \rho\s}}
\e_{\mn} \e_{\rho\s},
\end{align}
where $\Sigma$ is the $(D-2)$-dimensional surface of the event horizon,
${\cal L}$ is the Lagrangian density, and
$\e_\mn$ denotes the volume element binormal to $\Sigma$.
Although it was originally proposed for asymptotically flat spacetime,
we assume that it is applicable in the present system.
In \cite{Dutta}, it is confirmed that Wald's formula for asymptotically
AdS black hole solutions is still valid for gravity theories
with higher-curvature corrections, by comparing the result with the one obtained
by the Euclidean regularization method.
For the effective action in the string frame,
it gives
\begin{align}
S &= \frac{e^{-2 \phi_H} A_H }{4} \lh 1 +\frac{2\ta_2}{r_H^2}
\rh,
\end{align}
where $ A_H = A_{D-2} r_H^{D-2}$ is the area of the event horizon.
The entropy for the EGB gravity is given by
\begin{align}
S_{\rm (EGB)} &= \frac{A_H}{4} \lh 1 +\frac{2 \ta_2}{r_H^2} \rh.
\end{align}


\begin{figure}[h]
\begin{minipage}{50mm}
\begin{center}
\includegraphics[width=50mm]{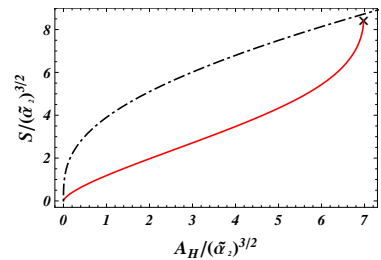} \\
(a) $A_H$-$S$ relation in five dimensions\\
\end{center}
\vspace{-4mm}
\begin{center}
\includegraphics[width=50mm]{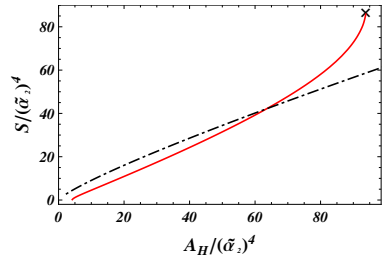} \\
(b) $A_H$-$S$ relation in ten dimensions\\
\end{center}
\end{minipage}
\caption{The relations between the horizon area and
entropy (a) in five dimensions and (b) in ten dimensions for $k=-1$.
The solid line in red and the dashed line in black correspond
to the cases of the DEGB and EGB gravity theories, respectively.
The maximum horizon area in five dimensions is given by
$A_{H, \rm{max}}=
6.9782851\, \ta_2^{3/2}$ denoted by $\times$,
which corresponds to $r_H = 0.70708722\, \ta_2^{1/2}$.
In ten dimensions the range of the horizon area of the obtained
numerical black hole solutions
is given by $4.1913928\, \ta_2^{4}< A_{H}
< 93.8048858\, \ta_2^{4}$, which corresponds
to the range of horizon radii as $ 0.78293332 \, \ta_2^{1/2}< r_H
< 1.1546707\, \ta_2^{1/2}$.
}
\label{AvsS}
\end{figure}

We depict the area-entropy relation in five and ten dimensions in Fig.~\ref{AvsS}.
Since we have the Gauss-Bonnet higher-curvature term,
the entropy is not proportional to the area; i.e.,
the Bekenstein-Hawking entropy formula is modified,
but it still increases monotonically as the horizon area increases.
We also plot the relation for the black holes
in the EGB gravity theory as a reference.
In any dimensions, the entropy increases as the horizon gets large.
The entropy approaches zero as the horizon approaches the smallest size,
which is zero for five dimensions and a finite size for other dimensions.
Around the largest horizon limit, the derivative of the entropy seems to blow up to infinity.
Since the entropy is smaller than that of EGB model
in the small horizon region, the entropy curve for the obtained numerical solution
crosses that for the EGB gravity in dimensions higher than five.
In five dimensions, it seems to reach numerically the lowest
bound before crossing.

\begin{figure}[h]
\begin{minipage}{50mm}
\begin{center}
\includegraphics[width=50mm]{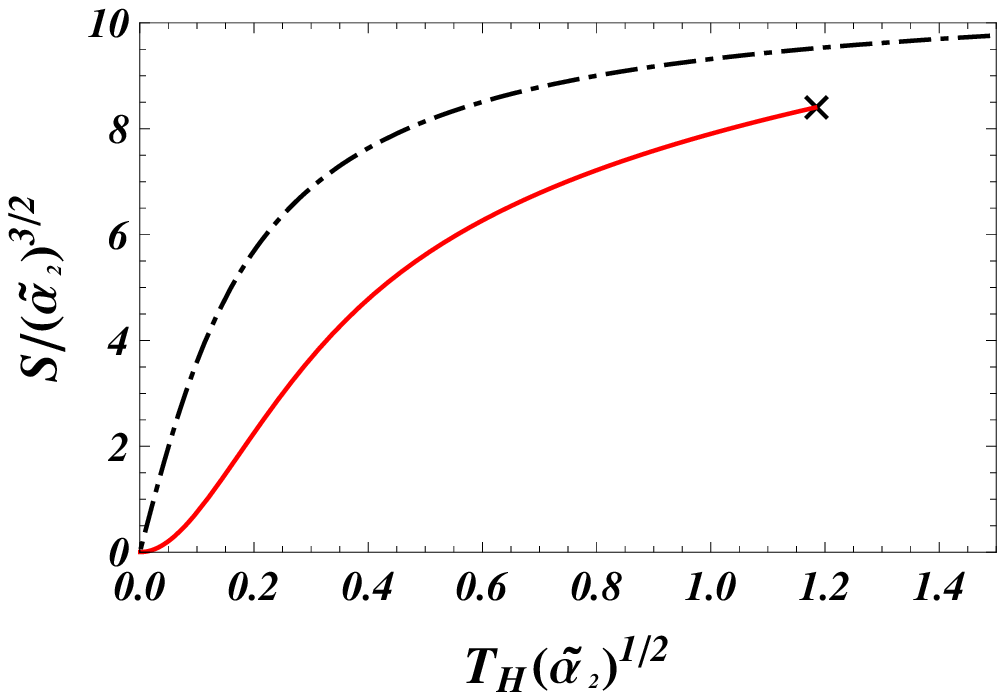} \\
(a) $T_H$-$S$ relation in five dimensions \\
\end{center}
\vspace{-4mm}
\begin{center}
\includegraphics[width=50mm]{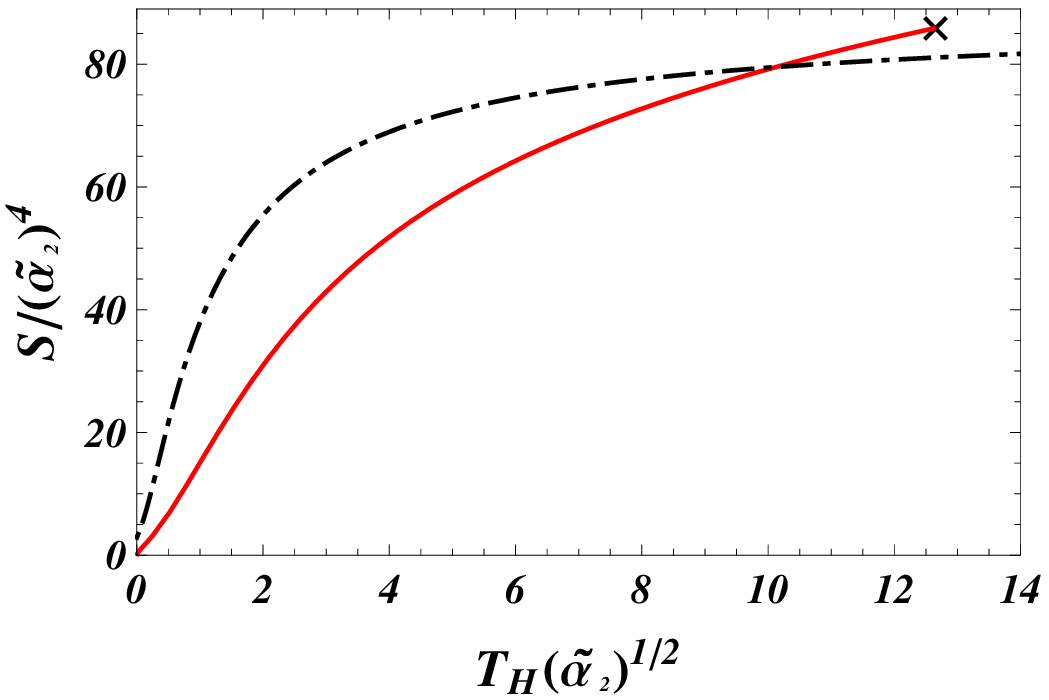} \\
(b) $T_H$-$S$ relation in ten dimensions
\end{center}
\end{minipage}
\caption{The relations between the temperature and entropy
(a) in five dimensions and (b) in ten dimensions for $k=-1$.
The solid line in red and the dot-dashed line in black correspond
to the cases of the DEGB and EGB models, respectively.
The maximum temperature is given by
$ T_{H, \rm{max}} = 1.18263054 \, \ta_2^{1/2} $
in five dimensions,
while $ T_{H, \rm{max}} = 12.6217055\, \ta_2^{1/2} $
 in ten dimensions.
Those end points are denoted by $\times$.
The temperature in the AdS branch of the EGB model
has no bound in any dimensions.
}
\label{TvsS}
\end{figure}

We also depict the temperature-entropy relation
in Fig.~\ref{TvsS}.
The temperature of the black hole vanishes
as horizon area approaches the minimum size in any dimension,
while it diverges as the horizon area approaches the
 maximum size.
This is similar to the case of the EGB gravity model.

\begin{figure}[h]
\begin{minipage}{50mm}
\begin{center}
\includegraphics[width=50mm]{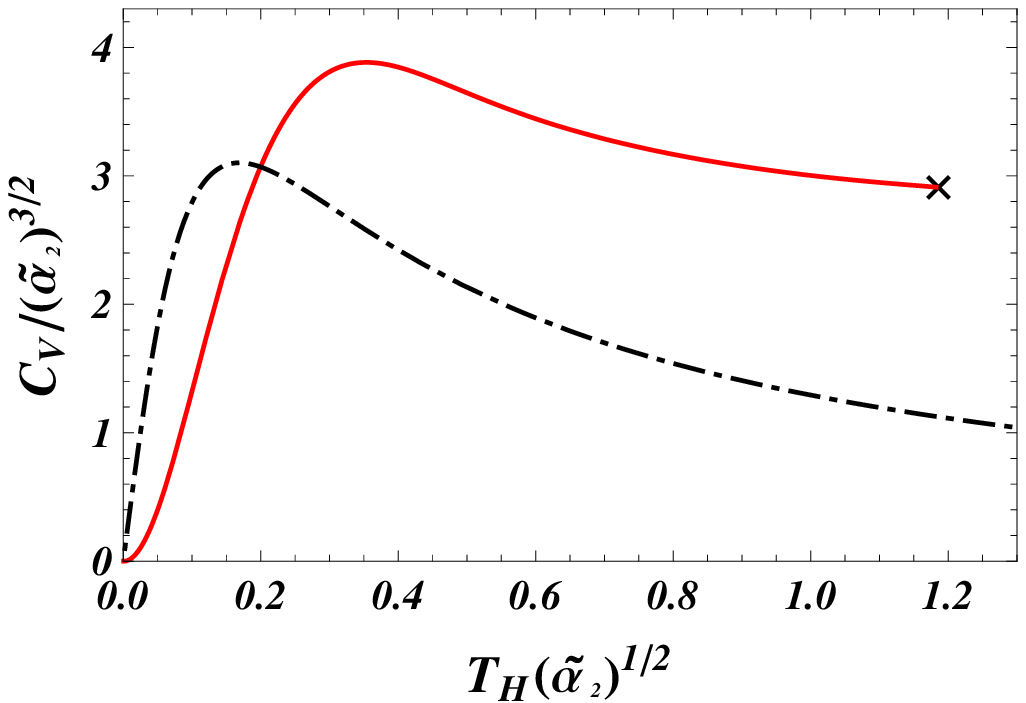} \\
(a) $T_H$-$C_{V}$ relation in five dimensions \\
\end{center}
\vspace{-4mm}
\begin{center}
\includegraphics[width=50mm]{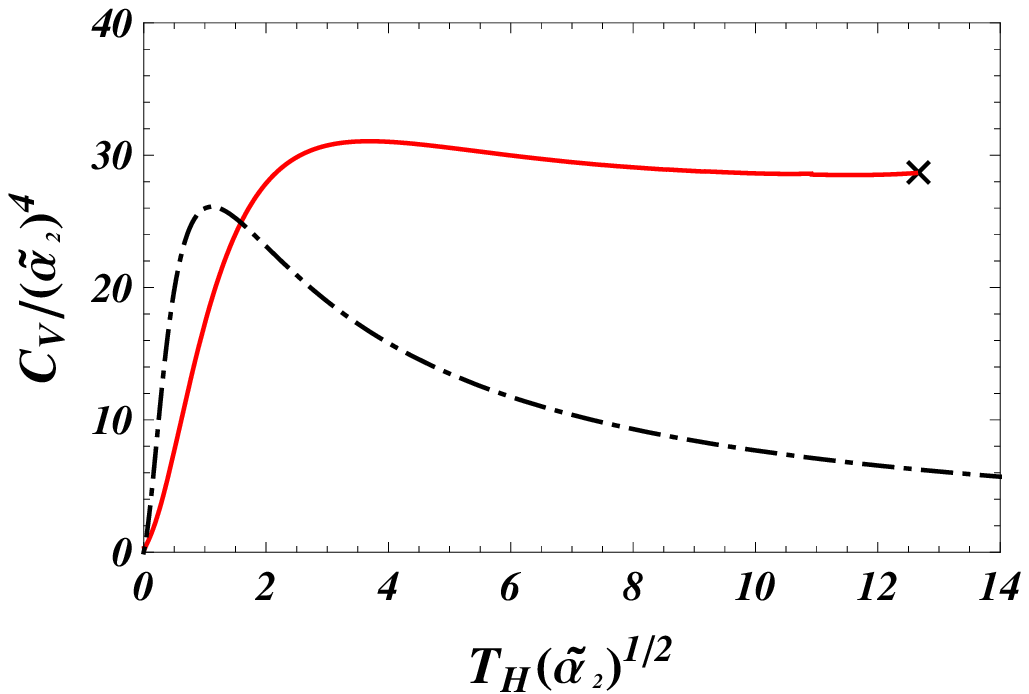} \\
(b) $T_H$-$C_{V}$ relation in ten dimensions
\end{center}
\end{minipage}
\caption{The relations between specific heat and temperature
(a) in five dimensions and (b) in ten dimensions for $k=-1$.
The solid line in red and the dot-dashed line in black correspond
to the cases of the DEGB and EGB models, respectively.
The maximum-temperature points are denoted by $\times$.
}
\label{CVvsT}
\end{figure}

The specific heat $C_V$ of a black hole solution is given by
\begin{align}
C_V = T \left( \frac{\pa S}{\pa T} \right)_V .
\end{align}
We show the relation between the specific heat and the
temperature in Fig.~\ref{CVvsT}.
The specific heat is positive for our solutions similar to that of AdS
black hole in the EGB model \cite{Neupane}.
In the nondilatonic EGB gravity, the specific heat is
negative for the $k=1$ black hole including the Schwarzschild solution,
while it is positive for $k=-1$ \cite{BH_EGB,AdSBH_EGB}.
Since the specific heat is always positive for our solution as shown
in Fig.~\ref{CVvsT},
we do not expect the Hawking-Page transition.


In order to discuss black hole thermodynamics, we have to
find a mass and a scalar charge of a black hole.
To extract the terms of a ``mass" and a ``scalar charge",
 we introduce the following variables as
\begin{align}
m[r] &\equiv r^{D-3-2p}\left(f[r] -\bar f[r]\right)
\label{subt_f} \\
&\rightarrow \mu_f~~~{\rm as}~~~r\rightarrow \infty,
\nonumber
\\[.5em]
\Phi[r] &\equiv r^{D-1-2p}\left(\phi[r]
 - \bar\phi[r]
\right)
\label{subt_phi}\\
&
\rightarrow \mu_\phi~~~{\rm as}~~~r\rightarrow \infty
\nonumber
\,.
\end{align}

\begin{figure}[h]
\begin{minipage}{50mm}
\begin{center}
\includegraphics[width=50mm]{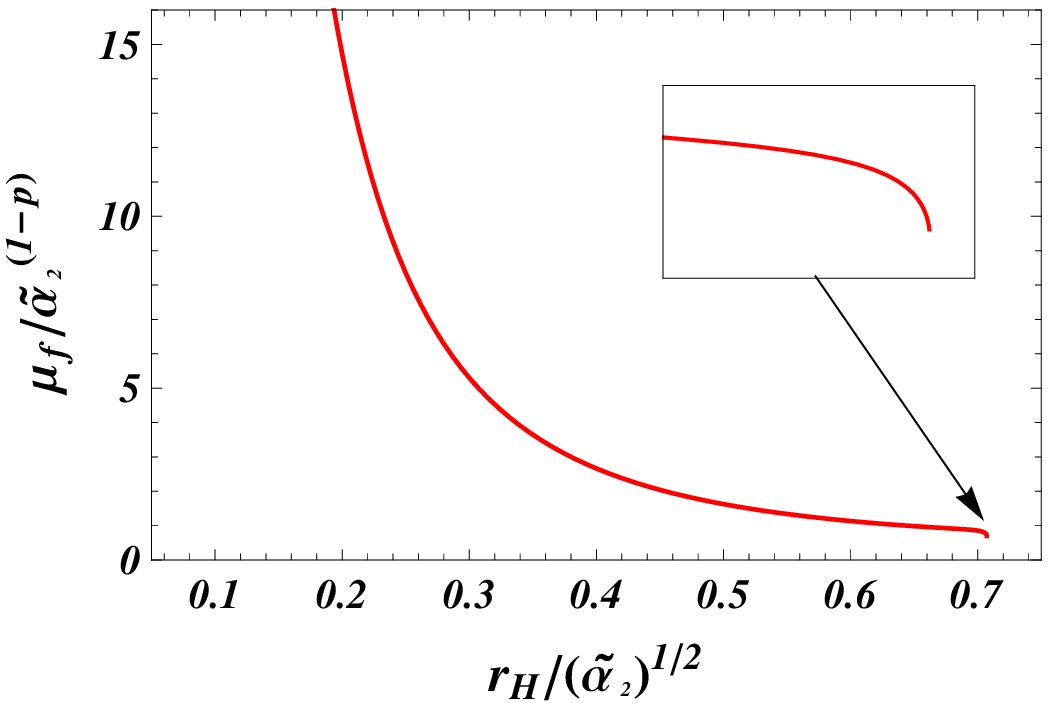}\\
(a) "mass" in five dimensions
\end{center}
\vspace{-3mm}
\begin{center}
\includegraphics[width=50mm]{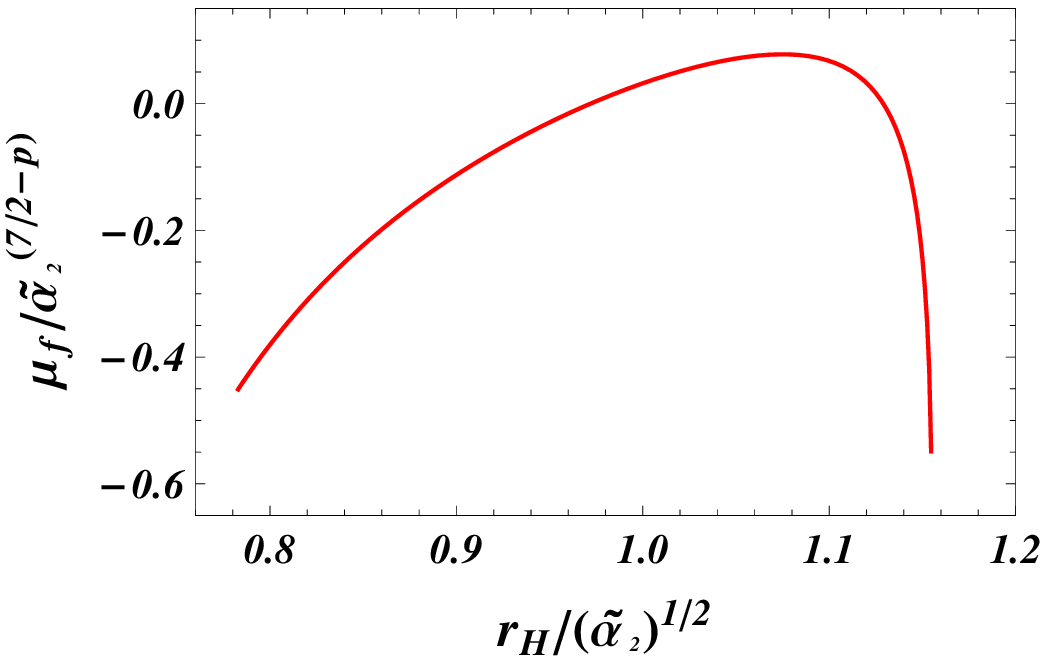}\\
(b) "mass" in ten dimensions
\end{center}
\end{minipage}
\caption{``Mass" $\mu_f$ in terms of the horizon radius for $k=-1$.
}
\label{mass_rh_5_10}
\end{figure}

\begin{figure}[h]
\begin{minipage}{50mm}
\begin{center}
\includegraphics[width=50mm]{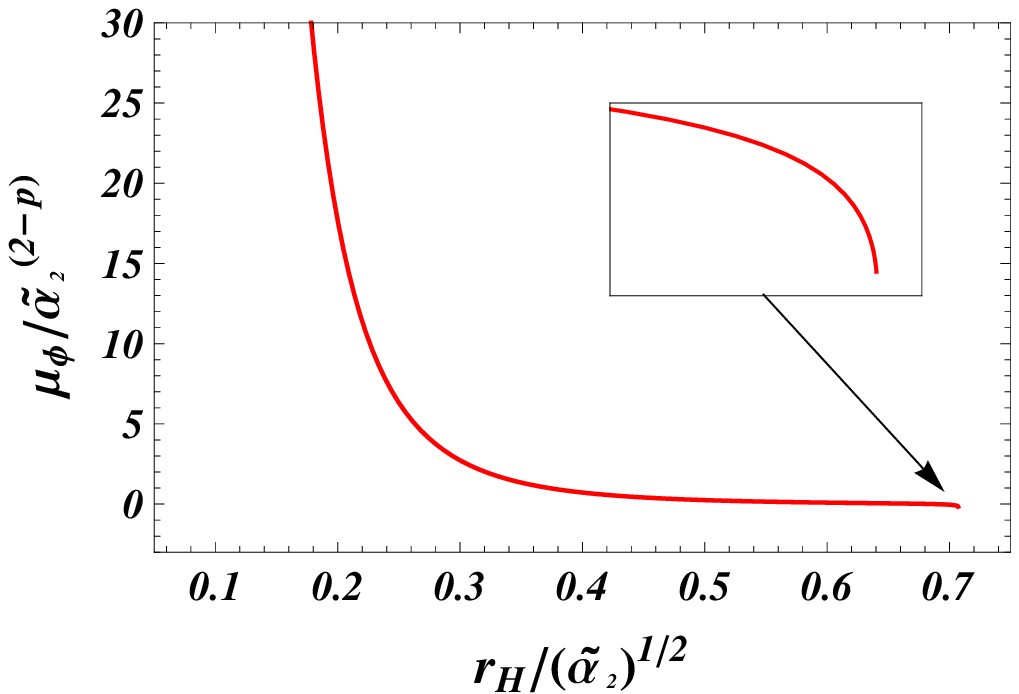}\\
(a) ``scalar charge" in five dimensions
\end{center}
\vspace{-3mm}
\begin{center}
\includegraphics[width=50mm]{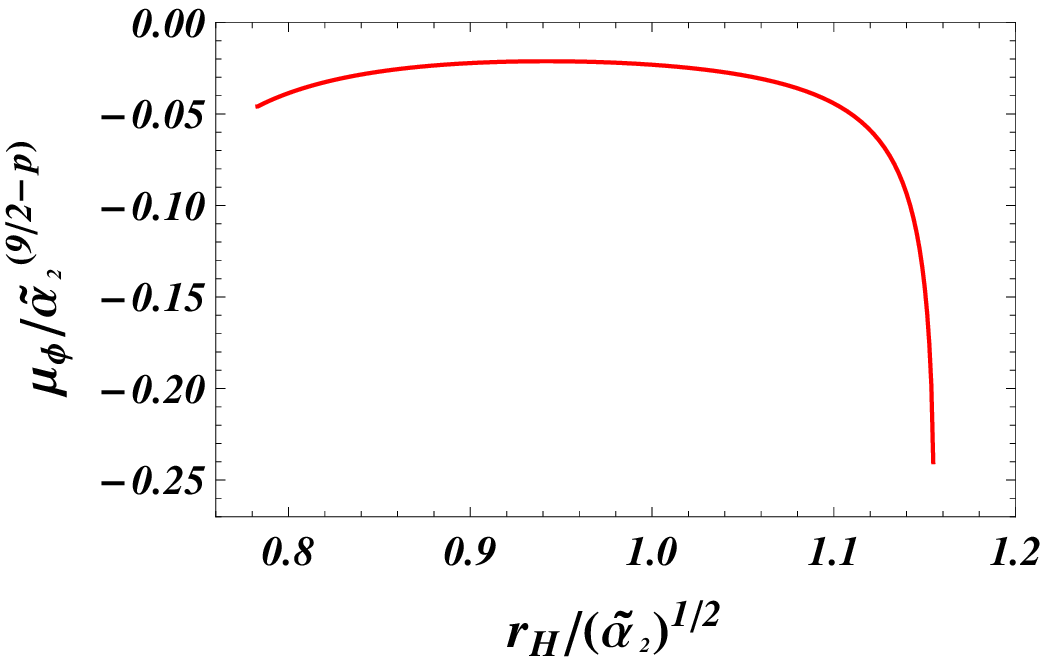}\\
(b) ``scalar charge" in ten dimensions
\end{center}
\end{minipage}
\caption{The ``scalar charge" $\mu_\phi$ in terms of the horizon radius for $k=-1$.}
\label{phi_rh_5_10}
\end{figure}

In Figs.~\ref{mass_rh_5_10} and \ref{phi_rh_5_10}, we depict
``mass" and ``scalar charge" defined by \eq{subt_f}
and \eq{subt_phi}, respectively.
The solutions are classified into two categories;
(a) odd dimensions and (b) even dimensions,
according to their behaviour of the ``mass'' and ``scalar charge".
In odd dimensions the ``mass'' monotonically decreases as the horizon radius
increases and approaches zero in the upper limit of the horizon radius.
On the other hand, the ``mass'' in even dimensions is not a monotonic function;
i.e., as the horizon radius increases, it first increases and then decreases just
after it reaches the maximum ``mass". It is not even positive.
The ``scalar charge" behaves similarly to the ``mass" function
in both odd and even dimensions.
However, the present ``mass" and ``scalar charge" are not
the physical ones.
Hence we cannot discuss thermodynamics
of our black holes furthermore.

These odd behaviours of the ``mass'' and ``scalar charge"
may be related to the fact that the AdS (or asymptotically AdS) spacetime of
our system is realized not by a negative cosmological constant
but by a dilaton field which diverges logarithmically at infinity.
This type of solution does not exist if we assume that a dilaton field is finite,
which is also true for the EDEGB case \cite{ohta_torii2}.

\section{Summary and Conclusion}
In the DEGB gravity,
we have obtained the exact AdS spacetimes with a planar symmetry
and constructed the asymptotically AdS black holes numerically.
These solutions are the dilatonic generalization of AdS spacetime
and the AdS branch of black hole solutions in
the EGB gravity.
A dilaton field, which diverges logarithmically at infinity,
plays the role of a negative cosmological constant.

The allowed horizon radii for asymptotically AdS black holes
seem to be inherited from those of the corresponding solutions
in the AdS branch in the EGB gravity.

As for the thermodynamical properties of the black holes,
the entropy and temperature are well defined and
their behaviours are similar to those of
the ``AdS branch" solution with a hyperbolic horizon in the EGB gravity.
However, we could not define the proper mass and charge,
although we have introduced a ``mass'' term and ``scalar charge'',
which may carry some physical properties of the black holes.
It makes it difficult for us to continue our discussion on thermodynamics such as
the first law further.
This strange behaviour of ``mass" and ``charge" may come from the fact that
a dilaton field diverges logarithmically at infinity.

We hope that AdS spacetimes and the asymptotically AdS black holes
obtained here give insight into the the AdS/CFT correspondence and
reveal aspects of strong coupling physics such as those studied in \cite{Cai1}.
In Appendix~A, we show that AdS spacetime is also a solution
for the models with more corrections in the string theory.
Examining the CFT duals of AdS spacetimes we obtained in Sec.~II and Appendix~A
would be an interesting issue.

\section*{Acknowledgments}

This work was supported in part by the Grant-in-Aid for
Scientific Research Fund of the JSPS (C) No. 20540283, No. 22540291,
No. 21$\cdot$09225, and
(A) No. 22244030.

\appendix
\begin{widetext}

\section{AdS spacetime in dilatonic Lovelock gravity}
\label{AdS_Lovelock}

In this appendix, we extend our AdS solution for $k=0$ in the DEGB
theory to the $D$-dimensional
Lovelock gravity coupled to a dilaton field, whose action in the string frame is given by
\begin{align}
S_{\!S} =\frac{1}{2\k_{D}^2} \int d^D \! x \sqrt{-g} \ e^{-2\phi}
 \, \biggl( R[g] +4{\na}_\mu \phi {\na}^\mu \phi
 + \sum^{[D/2]}_{n=2} \a_{n} E_{2n}[g] \biggr) ,
\end{align}
where $\a_n$ is the $n$th order coupling constant and
the $n$th order Euler density $E_{2n}[g]$
is defined by
\begin{align}
E_{2n}[g]
=\frac{-1}{2^n \, (D-2n)!}
 R^{\mu_1\nu_1}\!_{\rho_1\sigma_1}[g]
 \cdots R^{\mu_{n}\nu_{n}}\!_{\rho_{n}\sigma_{n}}[g]
\, \epsilon^{\a_1\a_2 \cdots\a_{D-2n}
\rho_1\sigma_1 \cdots\rho_{n}\sigma_{n}}\epsilon_{\a_1\a_2\cdots
\a_{D-2n}\mu_1\nu_1\cdots\mu_{n}\nu_{n}}.\nonumber
\end{align}
This special combination of curvatures guarantees that
$E_{2n}[g]$ is totally divergent in a $2n$-dimensional space, and
its integration gives just a topological invariant.
For $n = 2$, it is known as the Gauss-Bonnet curvature term,
which we have analyzed in the text.

In order to find an AdS spacetime, we adopt the same metric ansatz as in Sec.~II.
The explicit form of the action with this ansatz is given by
\begin{align}
S_{\!S}=
\frac{1}{2\k_{D}^2} \int d^D \!x e^{ {W}-2\phi}
 \biggl[ & e^{-2\lambda} \bigl( 2(D-2) {Y} +(D-2)_3 {A}
-4 \phi' \nu' \bigr) \biggr.
 +4 e^{-2\lambda} \phi'^2 \nn
 + & \sum^{[D/2]}_{n=2} \ta_{n} e^{-2n\lambda}
 \bigl( (D-2n)_{2n+1} {A}^{n} +2n (D-2n) {Y} {A}^{n-1}
 -4 n \phi' \nu' {A}^{n-1}
\bigr) \biggl. \biggr] ,
\end{align}
where we have dropped surface terms and used a normalized coupling constant
$\ta_{n} \equiv (D-2)_{2n-1} \a_n$.
The metric components $\lambda, \mu, $ and
$\nu$ are defined by the metric ansatz
(\ref{metric_string}), and
$W, Y,$ and $A$ are given by Eq. (\ref{YAW}).
Taking the variation of this action with respect to
$\nu$, $\lambda$, $\mu$, and $\phi$, we find four basic equations:
\begin{align}
F_{{\rm (L)}\phi}:=& -2\Bigl\{ (D-2)_{3} A +2 (D-2) (Y+Z) +2 X
+4\bigl[ \lh \nu' -\la' +(D-2)\mu' -\phi' \rh \phi' +\phi'' \bigr]
\label{eq_phi_Love} \nn
 +&\sum_n^{[D/2]} \ta_{n} e^{-2n\la} \Bigl[
(D-2n)_{2n+1} A^{n} +2n (D-2n) (Y+Z) A^{n-1}
 +2n X A^{n-1} +4 n_1 Y Z A^{n-2}
 \Bigr] \Bigr\} =0,
\end{align}
\begin{align}
F_{{\rm (L)}\nu}:=&
(D-2)_3 A +2(D-2)(Y + 2 \phi' \mu') +4( \phi'' -\phi'^2 -\phi' \la' )
\label{eq_nu_Love} \nn
+&\sum_n^{[D/2]} \ta_{n} e^{-2n\la} \Bigl[
(D-2n)_{2n+1} A^{n} +2n (D-2n) \lh Y + 2 \phi' \mu' \rh A^{n-1} \nn
& \hspace{20mm}
+4 n \lh \phi_n'' -2 \phi'^2 \rh A^{n-1} -4 n \phi' \la' A^{n-1}
 +8 n_1 \phi' \mu' Y A^{n-2}
 \Bigr]=0,~~~~~~~~~~~~~~~~~~~~~~~~~~~~~~~~~~~
\end{align}
\begin{align}
F_{{\rm (L)}\lambda}:=&
(D-2)_3 A +2(D-2)(Z +2 \phi' \mu') +4 (\nu' -\phi') \phi'
\label{eq_la_Love} \nn
+&\sum_n^{[D/2]} \ta_{n} e^{-2n\la} \Bigl[
(D-2n)_{2n+1} A^{n} +2n (D-2n) \lh Z +2 \phi' \mu' \rh A^{n-1}
+4 n \phi' \nu' A^{n-1} +8 n_1 \phi' \mu' Z A^{n-2}
 \Bigr]
=0,
\end{align}
\begin{align}
F_{{\rm (L)}\mu}:=&
(D-2)_4 A +2(D-2)_3 (Y +Z +2\phi'\mu') +2 (D-2) X
+4(D-2) \lh \phi'' - \phi'^2 +\phi' \nu' -\phi' \la' \rh
\label{eq_mu_Love} \nn
+&\sum_n^{[D/2]} \ta_{n} e^{-2n\la} \Bigl[
 (D-2n)_{2n+2} A^{n} +2n (D-2n)_{2n+1} ( Y +Z + 2 \phi' \mu' ) A^{n-1}
+4 n (D-2n) ( \phi'' -2 \phi'^2 ) A^{n-1} \nn
& \hspace{17mm} +2n (D-2n) X A^{n-1} +4 n (D-2n) ( \nu' - \la' ) \phi' A^{n-1}
+8 n_1 (D-2n) ( Y + Z ) \phi' \mu' A^{n-2} \nn
& \hspace{17mm} +4 n_1 (D-2n) Y Z A^{n-2}
 +8 n_1 \lh \phi'' -2 \phi'^2 \rh Z A^{n-2} \nn
& \hspace{17mm} +8 n_1 \lh \mu' X + \nu' Y -\la' Z \rh \phi' A^{n-2}
 +16 n_2 \phi'\mu' Y Z A^{n-3} \Bigr]
=0,
\end{align}
\end{widetext}
where we have introduced
\begin{eqnarray}
&{X}=-\left( \nu'' +\nu'^2 -\lambda' \nu' \right) \,,~
{Z}=-\mu' \nu'
\,,
\end{eqnarray}
in addition to \eq{YAW}.
The Bianchi identity is also satisfied.
Hence we have to solve three equations as in Sec.~II.

In order to obtain an exact AdS spacetime, we take the same
ansatz (\ref{ansatz_k0}) and (\ref{ansatz_k1}) as in Sec.~II.
We then find the following three algebraic equations:
\begin{eqnarray}
&&\left[D_1 -4 p(D-1) +4 p^2\right]
+D_1 {\rm S}_1 =0
\,,
\label{eq_AdS_phi}
 \\
&&p+p(1+2p) {\rm S}_2=0
 \,,
\label{eq_AdS_nu_la}
\\
&&L\left[(D-1)_2-4 p(D-2) +4 p^2\right]
+{\rm S}_3=0
\,,~~~
\label{eq_AdS_nu}
\end{eqnarray}
where $L\equiv -1/\ell^2$ and
\begin{align}
{\rm S}_1 &= \!\sum_{n=2}^{[D/2]}\!\ta_{n} L^{n-1}  \, ,\nn
{\rm S}_2 &= \!\sum_{n=2}^{[D/2]}\!\ta_{n} L^{n-1}n \, ,\\
{\rm S}_3 &= \!\sum_{n=2}^{[D/2]}\!\ta_{n}
             L^n \!\left[ (D-1)(D-2n) -4 n p(D-2) +8 n p^2 \right],
\nonumber
\end{align}
 Equation (\ref{eq_AdS_phi}) with Eq. (\ref{eq_AdS_nu_la}) gives Eq.
(\ref{eq_AdS_nu}).
It means that three algebraic equations are not independent.
Hence we have to solve the following two algebraic equations:
(\ref{eq_AdS_phi}) and (\ref{eq_AdS_nu_la})
 for two variables ($\varpi=1+2p$ and $L=-1/\ell^2$), i.e.,
\begin{align}
&
1+\varpi~{\rm S}_2 =0,
\\
&
\varpi^2-2D\varpi+(D^2+D-1)+D_1~{\rm S}_1
=0.
\end{align}
Eliminating $\varpi$, we obtain
\begin{align}
F(L)\equiv {\rm S}_2^2 \left( D^2+D-1+D_1 ~{\rm S}_1 \right)
-2D~{\rm S}_2 +1=0.
\label{eq_L}
\end{align}
$F(L)$ is the $3(n_{\rm max}-1)$th order algebraic equation with respect to $L$,
where $n_{\rm max}$ is the highest order with nontrivial $\ta_{n}$
($n_{\rm max}\leq [D/2]$).
If $n_{\rm max}$ is even , i.e., $3(n_{\rm max}-1)$ is odd, we always find
a negative solution ($L<0$) for Eq. (\ref{eq_L}), since $F(0)=1>0$,
that is, an AdS space.
On the other hand, if $n_{\rm max}$ is odd, the existence of AdS space is
no longer guaranteed.

If we have only one Lovelock term, for example, 
the $n_0$th Lovelock term $(n_0\leq [D/2])$,
we can conclude more concretely
\begin{eqnarray}
&&
1+\varpi\ta_{n_0} L^{{n_0}-1} n_0 =0
\,,
\label{eq_xL1}
\\
&&
\varpi^2-2D\varpi+D^2+D-1+\ta_{n_0} L^{{n_0}-1} =0
\,.
~~~~~~~~~~
\label{eq_xL2}
\\
\nonumber
\end{eqnarray}

Eliminating the terms with $L^{{n_0}-1}$, we obtain the
third-order algebraic equation for $\varpi$:
\begin{eqnarray}
&&
\varpi^3-2D\varpi^2+(D^2+D-1)\varpi-\frac{D(D-1)}{n_0}=0
\,,\nonumber \\
&&~
\end{eqnarray}
which has always a positive real root $\varpi_0 (>0)$.
{}From (\ref{eq_xL1}) we find
\begin{eqnarray}
L_{*}^{n_0-1}=-\frac{1}{n_0 \ta_{n_0}\varpi_0}
\,.
\end{eqnarray}
Since $\ta_{n_0}>0$, we find a solution with $L_{*}<0$
if $n_0$ is even but no solution for odd $n_0$.

In the string theory, we expect only even $n$'s.
Hence we always find at least one AdS solution.
For example, we consider the case of the heterotic string model such as
($D=10$, $\alpha_2=\alpha'/8$, $\alpha_3=0$
and $\alpha_4=\alpha'^3/8$). Then we find
\begin{eqnarray}
p = -0.36469898\,,~ \ell = 3.91866401\alpha'^{1/2}
\,.
\end{eqnarray}

\section{``Solitonic" solution with AdS boundary}
\label{solitonic}
We also find a solitonic solution without a horizon
but with a regular center, which approaches
asymptotically an AdS spacetime.
In this appendix, we leave the global topology $k$ free at first
and fix it later.
To solve the basic equations under a regularity condition
at the origin ($r=0$),
we expand the variables around the origin $r=0$ as
\begin{eqnarray}
f(r) &=& \hat f_0 +\hat f_1 r +\hat f_2 r^2 +{\cal O}(r^3) ~,
\nn
\d(r) &=& \hat \d_0 +\hat \d_1 r +\hat \d_2 r^2 +{\cal O}(r^3) ~,
\label{expand_origin}
\\
\phi(r) &=& \hat \phi_0 +\hat \phi_1 r + \hat \phi_2 r^2 +{\cal O}(r^3) ~.
 \nonumber
\end{eqnarray}
The shift symmetry of $\phi$ and $\delta$
allows us to choose $\hat \d_0 = \hat \phi_0 =0$.
Substituting \eq{expand_origin} into the field equations,
we find that $\hat f_0=k$ in the leading equations and
$\hat f_1=\hat\d_1=\hat\phi_1=0$ in the second leading ones.
The third leading equations give
\begin{eqnarray}
&&D(\hat f_2-1)\hat f_2 +4 k (1-2\hat f_2) \hat\d_2
 +8 k \hat\phi_2 =0~,
\nn
&&D(\hat f_2-1)\hat f_2 +2(1-2\hat f_2)
(\hat f_2+4 k\hat\phi_2) =0 ~,
\label{quad_n}
\\
&&(1-2 \hat f_2) (2\hat \phi_2 -(D-2)\hat \d_2)
-4\hat \d_2 (\hat f_2 +4 k \hat \phi_2 ) =0 ~.
\nonumber
\end{eqnarray}
We show the numerical solutions of \eq{quad_n}
in Table \ref{table_5}.

\begin{table}[h]
\caption{The solutions for Eq.~\eq{quad_n} for $k=1$ which give
an asymptotically AdS spacetime.
}
\begin{center}
\begin{tabular}{|c|c|c|c|}
\hline
$D$& ~~~$\hat f_2$~~~ & ~~~$\hat \d_2$~~~ & ~~~$\hat \phi_2$~~~
\\
\hline
4&~~~~$1.0850432$~~~~&~~~~$-0.31739860$~~~~&~~~~$-0.23182971$~~~~
\\
\hline
5&~~~~$1.1231984$~~~~&~~~~$-0.20045991$~~~~&~~~~$-0.21141146$~~~~
\\
\hline
6&~~~~$1.1298841$~~~~&~~~~$-0.13500237$~~~~&~~~~$-0.19510138$~~~~
\\
\hline
7&~~~~$1.1241759$~~~~&~~~~$-0.097812087$~~~~&~~~~$-0.18319807$~~~~
\\
\hline
8&~~~~$1.1150320$~~~~&~~~~$-0.075149215$~~~~&~~~~$-0.17448348$~~~~
\\
\hline
9&~~~~$1.1056361$~~~~&~~~~$-0.060330293$~~~~&~~~~$-0.16793263$~~~~
\\
\hline
10&~~~~$1.0969813$~~~~&~~~~$-0.050055266$~~~~&~~~~$-0.16286547$~~~~
\\
\hline
\end{tabular}
\label{table_5}
\end{center}
\end{table}

\begin{figure}[h]
\includegraphics[width=50mm]{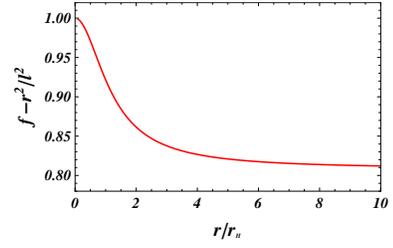}
\\
(a) $f(r)-r^2/\ell^2$ \\
\hskip 4cm
\includegraphics[width=50mm]{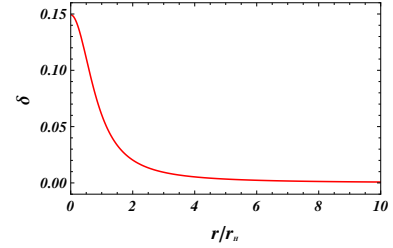}
\\
(b) $\delta(r)$ \\
\hskip 4cm
\includegraphics[width=50mm]{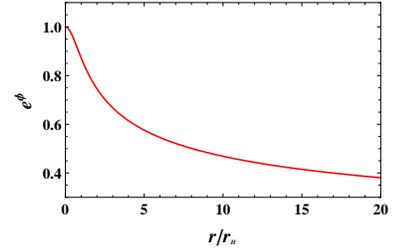}\\
(c) $e^{\phi(r)}$ \\
\caption{The solitonic AdS solution in five dimensions for $k=1$.
We depict the configurations of $[f(r)-r^2/\ell^2]$,
$\delta(r)$, and $e^{\phi(r)}$.
}
\label{soliton_AdS_5D}
\end{figure}

Adopting  \eq{expand_origin} with the solutions of
\eq{quad_n} as the boundary conditions
at the origin, we numerically solve the field equations.
Note that we have no free parameter at the origin.
We find two branches of solutions for $k=1$:
One approaches asymptotically an AdS spacetime,
and the other evolves into a singularity.
We depict our numerical solution in Fig.~\ref{soliton_AdS_5D}.

Analyzing the asymptotic behaviours, we find a 
similar result to the AdS black hole with $k=-1$
discussed in Sec.~\ref{boundary}.
Just as the black hole solutions, we present the ``mass''
and ``scalar charge''
in Table \ref{table_6}, which depend only on the spacetime dimensions $D$
and the coupling constant $\alpha_2$.
No additional free parameter appears.

For $k=-1$, for which we obtain the AdS black hole,
 we could not obtain a regular solution 
but always find a singularity at a finite radius.
For $k=0$, we obtain the exact AdS spacetime studied in Sec.~II.
\vspace*{1mm}
\begin{table}[h]
\caption{``Mass" and the ``scalar charge" introduced in \eq{subt_f} and \eq{subt_phi}
for $k=1$.}
\begin{center}
\begin{tabular}{|c|c|c|}
\hline
$D$& ~~~$ \mu_f $~~~ & ~~~$\mu_{\phi}$~~~
\\
\hline
4&~~~~$-0.64077734$~~~~&~~~~$-0.056993109$~~~~
\\
\hline
5&~~~~$0.17163694$~~~~&~~~~$0.016651835$~~~~
\\
\hline
6&~~~~$0.13056579$~~~~&~~~~$0.013608420$~~~~
\\
\hline
7&~~~~$-0.067443667$~~~~&~~~~$-0.0074559764$~~~~
\\
\hline
8&~~~~$-0.061032657$~~~~&~~~~$-0.0070525786$~~~~
\\
\hline
9&~~~~$0.039077656$~~~~&~~~~$0.0046311897$~~~~
\\
\hline
10&~~~~$0.037257968$~~~~&~~~~$0.0045987537$~~~~
\\
\hline
\end{tabular}
\label{table_6}
\end{center}
\end{table}

\section{Cosmological solution}
\label{cosmological}
For $k=-1$,
we also find a cosmological solution instead of an AdS black hole.
When we integrate the basic equations with the boundary conditions
for a regular horizon, we find that
the metric function $F$ is always negative
 in the larger-radius branch in Table \ref{table_2}.
It means that the ``time" coordinate $t$ plays no longer a role of time.
Instead, the ``radial" coordinate $r$ becomes time.
Hence we exchange the characters of two coordinates
as $\eta:=r$ and $\xi:=t$.
The line element is given by
\begin{align}
ds^2 = - f_C^{-1}(\eta)
d\eta^2 +f_C(\eta)e^{-2\delta(\eta)}
d\xi^2
+ \eta^2 d \Sigma_{-1}^2
\,, \nonumber
\end{align}
where
$f_C(\eta)=-f(r)$.
The asymptotic behaviour of the metric functions and
the dilaton field is given by
\begin{align}
f_C[\eta] &= \left[ 1 - \lh \frac{\eta_{\infty}}{\eta}\rh^{D-3} \right]
 + {\cal O}\lh \frac{1}{\eta^{D-3}}\rh \\
\d [\eta] &= {\cal O}\lh \frac{1}{\eta^{2(D-3)}}\rh \\
\phi[\eta] &= \phi_{\infty} + {\cal O}\lh \frac{1}{\eta^{2(D-3)}}\rh
\end{align}
as $\eta\rightarrow \infty$,
where $\eta_{\infty}$ is a constant.
Near the ``horizon", $f_C(\eta)$ vanishes as
$f_C(\eta)\propto (\eta-\eta_H)$, where $\eta_H
:=r_{H}$
is the ``horizon" radius.

We introduce a cosmic time $\tau$ by
\begin{align}
\tau = \int_{\eta_H}^\eta d\eta \, f_C^{-1/2}(\eta) \ .
\end{align}
Using this cosmic time,
we can rewrite the line element as
\begin{align}
ds^2 = - d\tau^2 + a^2 ( \tau ) d \Sigma_{-1}^2
+ b^2( \tau ) d\xi^2
\,,
\end{align}
where the metric functions $a(\tau)$ and $b(\tau)$ are defined by
\begin{align}
a:=\eta(\tau)
\,,~~
b:= f_C^{1/2}e^{-\delta} \,,
\end{align}
and the asymptotic behaviours as $\tau \rightarrow \infty$
are
\begin{align}
a\rightarrow
\tau \lh 1 + \frac{\tau_{\infty}^{D-3}}{2(D-4) \tau^{D-3}} \rh
\,,
b \rightarrow
1 - \frac{\tau_{\infty}^{D-3}}{2\tau^{D-3}} \, ,~~
\end{align}
where $\tau_{\infty}:=\eta_{\infty}$.
The asymptotic metric is given by
\begin{align}
ds^2 = - d\tau^2 + \tau^2 d \Sigma_{-1}^2 + d \xi^2
\,,
\end{align}
which is a [$(D-1)$-dimensional Milne universe] $\times {\bf R}^1$.
See \cite{Milne} for a Milne universe.

\begin{figure}[h]
\begin{minipage}{70mm}
\begin{center}
\includegraphics[width=50mm]{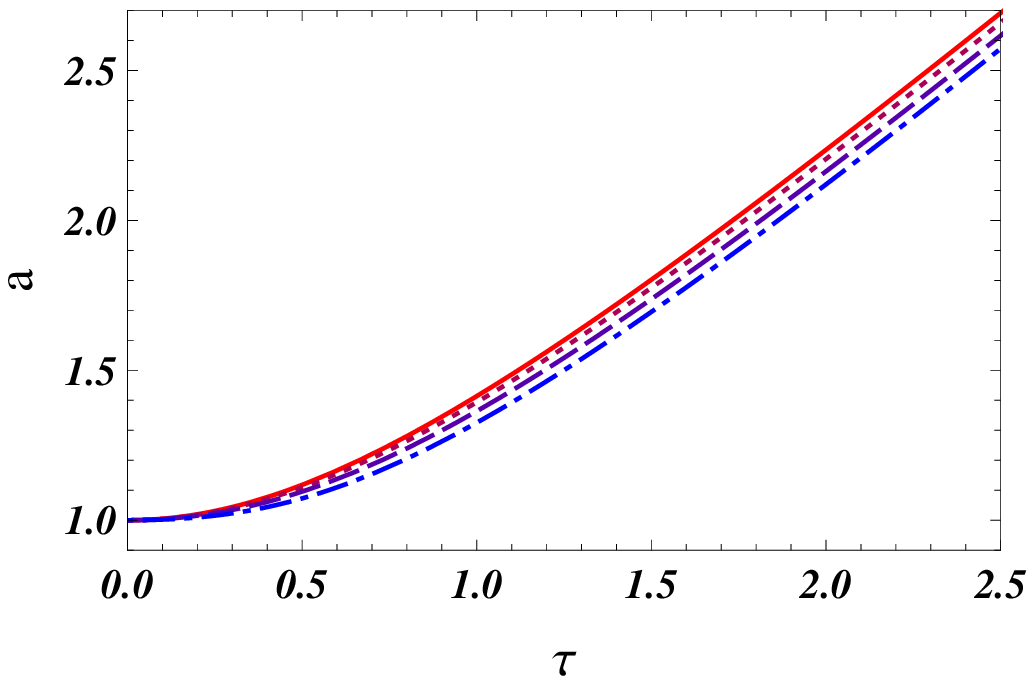}\\
(a) $a(\tau)[=\eta(\tau)]$
\end{center}
\end{minipage}
\begin{minipage}{70mm}
\begin{center}
\includegraphics[width=50mm]{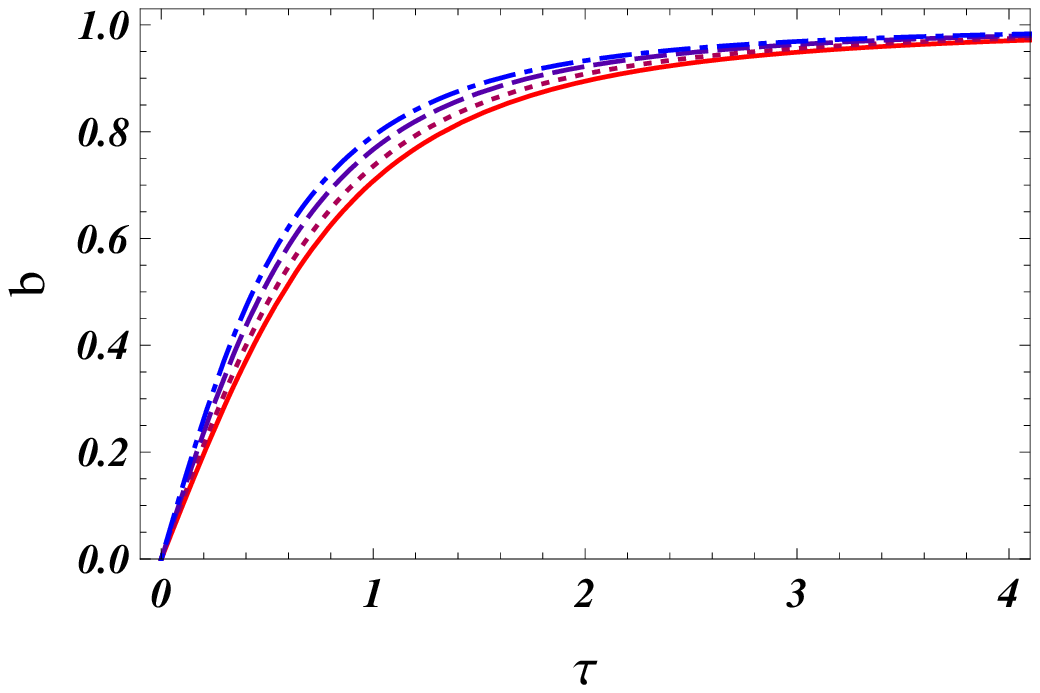}\\
(b) $b(\tau) $
\end{center}
\end{minipage}
\caption{We depict two scale factors $a(\tau)[=\eta(\tau)]$ and $b(\tau) $
in five dimensions for $k=-1$.
We choose the following four values for the ``horizon radius":
$\eta_H=\,$31.6227766\,$\ta_2^{1/2}$ (solid line), 4.14479601\,
$\ta_2^{1/2}$ (dotted line),
2.97463549\,$\ta_2^{1/2}$ (dashed line), and 2.62725036\,
$\ta_2^{1/2}$ (dot-dashed line).
}
\label{Cosmological}
\end{figure}

We depict numerical solutions in Fig.~\ref{Cosmological}.
It shows that the spacetime starts from a big bang singularity
and evolves into the [$(D-1)$-dimensional Milne universe] $\times {\bf R}^1$.
These solutions are obtained in the larger-radius branch of
$r_H(=\eta_H)$ in Table~\ref{table_2}.

For $D=5$,
we show our numerical result in Fig.~\ref{Cosmological}.
By compactifying the fifth direction ($\xi$),
we find a four-dimensional bouncing universe.
The three-dimensional space with the scale factor $a$
bounces at $\tau=0$ and evolves into a Milne universe.
Although the fifth dimension with the scale factor $b$
starts from an initial big bang singularity, it eventually
approaches a constant space as $\tau\rightarrow \infty$.

\end{document}